\documentclass[aps,pra,twocolumn,superscriptaddress,showpacs,preprintnumbers,am
smath,amssymb]{revtex4-1}
\usepackage{times}
\usepackage{color}
\input pdfcolor.tex
\usepackage{colortbl,amsthm,amsmath,amssymb,txfonts}
\usepackage{graphicx}
\usepackage{epsfig,color}
\usepackage[colorlinks,bookmarks=false,citecolor=blue,linkcolor=red,urlcolor=blue]{hyperref}

\begin{document}

\title{Thermal transitions, pseudogap behavior and BCS-BEC crossover in Fermi-Fermi mixtures}
\author{Madhuparna Karmakar}
\affiliation{Harish chandra research institute, Chhatnag road, Jhunsi,
 Allahabad-211019, India, and \\ Homi Bhabha National Institute,
 Training School Complex, Anushakti Nagar, Mumbai 400085, India.}
\affiliation{Department of Physics, Indian Institute of technology Madras, 
Chennai-600036, India.}

\begin{abstract}
We study the mass imbalanced Fermi-Fermi mixture within the 
framework of a two-dimensional lattice fermion model. 
Based on the thermodynamic and species dependent quasiparticle 
behavior we map out the finite temperature phase diagram of this system 
and show that unlike the balanced Fermi superfluid there are now 
two different pseudogap regimes as PG-I and PG-II. While within the 
PG-I regime both the fermionic species are pseudogapped, PG-II 
corresponds to the regime where pseudogap feature survives only in the 
light species.
We believe that the single particle spectral features that we 
discuss in this paper are observable through the species resolved 
radio frequency spectroscopy and momentum resolved photo emission spectroscopy 
measurements on systems such as, 6$_{Li}$-40$_{K}$ mixture. 
We further investigate the interplay between the population and 
mass imbalances and report that at a fixed population imbalance 
the BCS-BEC crossover in a Fermi-Fermi  mixture would require a 
critical interaction (U$_{c}$), for the realization of the uniform
superfluid state. The effect of imbalance in mass on the exotic 
Fulde-Ferrell-Larkin-Ovchinnikov (FFLO) superfluid phase has been 
probed in detail in terms of the thermodynamic and quasiparticle
behavior of this phase. It has been observed that in spite of the s-wave 
symmetry of the pairing field a nodal superfluid gap is realized in the 
LO regime.
Our results on the various thermal scales and 
regimes are expected to serve as benchmarks for the experimental 
observations on 6$_{Li}$-40$_{K}$ mixture.
\end{abstract}

\date{\today}
\maketitle

\section{Introduction}

Ultracold atomic gases with the tunability of their interaction 
strength has proved to be a suitable quantum simulator for several 
many body phenomena. A principal one being the realization
of exotic superfluid phases in Fermi gases \cite{regal2004, zwierlein2004, kinast2004, bart2004}.
The experimental realization of the same continues to be illusive 
so far but that has not prevented the theoretical investigation of the 
various possibilities viz. p-wave superfluid \cite{regal2003, zhang2004, schunck2005,
ohashi2005, ho2005, gurarie2005, levinsen2007, iskin2005, inotani2012}, imbalanced 
superfluid \cite{sarma1963, liu2003, forbes2005, sheehy2007},
superfluid with hetero Cooper pairs \cite{wille2008, taglieber2008, voigt2009, costa2010, 
naik2010, spiegel2010, tiecke2010, lin2006, wu2006, iskin2006, iskin2007, 
pao2007, parish2007,orso2008, gezerlis2009, deiner2010, takemori2012, baarsma2010, 
baarsma2012, lan2013, ohashi2013} and Fermi superfluid
with spin-orbit interaction \cite{lin2011, wang2012, cheuk2012, jiang2011}. 

Among the various possibilities, 
imbalanced Fermi superfluids is one  
which has been widely explored.
Imbalance in Fermi superfluids can be realized through, 
(i) population imbalance or (ii) mass imbalance.
While the former has been investigated in detail both experimentally 
\cite{partridge2006, shin2008, schunck2007, shin2006, liao2010} and theoretically
\cite{torma2007, loh2010, scalleter2012, mpk2016, mpk2016_epjd},
studies carried out on unequal mass Fermi-Fermi mixtures are 
relatively few \cite{carlson2009, drut2015, roscher2014,stoof2010, ohashi2013, ohashi2014,levin_mass, batrouni2009}.
Experimentally a mass imbalanced Fermi-Fermi mixture
is achievable in a ${6}_{Li}-40_K$ mixture. 
While superfluidity in such a system is yet to be attained in 
experiments, the Fermi degenerate regime \cite{taglieber2008, naik2010}  
as well as the Feshbach resonance between $6_{Li}$ and $40_{K}$ atoms
\cite{wille2008, costa2010, naik2010} and formation of $6_{Li}$-$40_{K}$ 
heteromolecules \cite{voigt2009} are already a reality.
Furthermore, experimental realization of mixtures of other fermion species
(such as $161_{Dy}$, $163_{Dy}$, $167_{Er}$) are expected in future
\cite{lev2012, frisch2013}.

An experimentally addressable aspect of the mass imbalanced
mixture is its finite temperature behavior. It has been reported that 
for a double degenerate $6_{Li}-40_{K}$ mixture the Fermi temperatures 
are T$_{F}$$^{Li} = 390$nK and T$_{F}$$^{K} = 135$nK, for Li and K species,  
respectively \cite{grimm2010}. In comparison, for a balanced Fermi gas of 
$6_{Li}$, the Fermi temperature is known to 
be $T_{F} = 1.0\mu$K \cite{grimm2011} with the corresponding T$_{c}$ 
scale being T$_{c} \sim 0.15$T$_{F}$ \cite{solomon2010}. While it is
evident that in case of the mass imbalanced Fermi-Fermi mixture 
the thermal scales are significantly suppressed, the qualitative 
and quantitative behavior of the same is hitherto unknown. 

Keeping in pace with the experiments, efforts have been 
put in to theoretically investigate the behavior of mass 
imbalanced Fermi-Fermi mixtures within the framework of 
continuum models.
Density functional theory combined with local density approximation
\cite{roscher2014}, functional renormalization group studies etc. have 
been carried out
on mass imbalanced Fermi mixture at unitarity \cite{drut2015}.
The study involved inclusion of fluctuations beyond the mean field
and predicted the possibility of inhomogeneous superfluid state.
The problem has also been investigated using 
mean field theory (MFT) taking into account the effects of 
gaussian fluctuations \cite{stoof_prl2009, stoof2010}. The authors 
mapped out the polarization-temperature phase diagram at different 
mass as well as population imbalances at and away from unitarity.
It was shown that while for a mass balanced system, instability towards 
a supersolid phase accompanied by a Lifsitz point is observed only at 
weak interactions, the imbalance in mass promotes this behavior and 
makes it observable even at unitarity.  Among the other techniques 
T-matrix and extended T-matrix approaches \cite{ohashi2013, 
ohashi2014, levin_mass} are utilized 
to determine the thermal scales of the mass imbalanced mixture, both in
terms of its thermodynamic as well as quasiparticle behavior. 
It was observed that unlike the balanced Fermi gas the Fermi-Fermi 
mixture with imbalance in mass consists of more than one pseudogap scales 
\cite{ohashi2013, ohashi2014}.

Interestingly, within the framework of a lattice fermion model 
most of the theoretical investigations on imbalanced Fermi gases are carried out on 
systems with imbalance in population, through
improvised numerical and analytic techniques
 \cite{torma2007, loh2010, scalleter2012, torma2013, torma2014, zhang, mpk2016, mpk2016_epjd}. 
For the mass imbalanced mixture there are only few attempts 
that has been made within the framework of lattice Fermion model.
For example, the ground state behavior of one dimensional mass imbalanced 
system has been studied through Quantum Monte carlo (QMC) calculations \cite{batrouni2009}.
Recently a nonperturbative lattice Monte Carlo calculation 
was carried out to address the ground state of Fermi-Fermi 
mixture in two dimensions (2D) \cite{roscher2015}.
The study revealed that a mean field approach to the problem
grossly overestimates the ground state energy of the system.
The effect is likely to be more severe at finite temperature 
where crucial amplitude and phase fluctuations are neglected 
within a mean field scheme. 

An estimate of the inadequacy of 
mean field approach to the problem at finite temperature
can be made from the fact that mean field theory overestimates
the T$_{c}$ scales by a factor of more than $4$ both in case of balanced
\cite{engelbrecht1993, zwerger2007, strinati2004, troyer2006, bulgac2007, akkineni2007}
as well as population imbalanced Fermi superfluids \cite{mpk2016}. 
This is a crucial observation owing to the fact that many of the predictions 
for imbalanced Fermi superfluids are being made based on the mean 
field theory. 

While there is now a consensus about the thermal behavior of balanced Fermi 
superfluid \cite{randeria_taylor, paiva2d},
the same can not be said about the Fermi-Fermi mixture. Within the purview
of lattice fermions,
there is certainly a void in our present understanding of such mixtures,
specially at finite temperatures. On the other hand, a lattice fermion model 
is a suitable choice keeping in view the optical lattice experiments that 
are carried out on ultracold Fermi gases.   
What is the T$_{c}$ scale of a mixture such as 6$_{Li}$-40$_{K}$?
How does such a system behave across the BCS-BEC crossover and most importantly
how does the pseudogap physics play out in the back ground of an 
imbalance of fermionic masses in the system? While an experimental 
investigation to answer such questions is awaited, one can certainly 
make theoretical predictions. 

Motivated by these questions, in this paper we present a detail finite 
temperature analysis of mass imbalanced Fermi mixture within the framework of 
lattice fermions. We use a numerical technique which takes into 
account the phase fluctuations of the ordering field and 
can access the thermal transitions with quantitative correctness.
Apart from investigating the behavior of mass imbalanced system across
the BCS-BEC crossover we also investigate the interplay of mass and 
population imbalances. Before making quantitative predictions about the 
6$_{Li}$-40$_{K}$ mixture we have discussed how the exotic Fulde-Ferrell-Larkin-Ovchinnikov (FFLO) 
superfluid phase reacts to the imbalance in mass in terms of 
thermodynamic and quasiparticle behavior. We highlight our principal 
observations below before proceeding to discuss the numerical technique 
and the results obtained from the same. 

\begin{itemize}
\item{Imbalance in mass leads to strong suppression of T$_{c}$
across the BCS-BEC crossover regime. Close to unitarity (U $\sim$ 4t$_{L}$)
for a mass balanced 6$_{Li}$ gas T$_{c}^{bal}$$\sim$0.15t$_{L}$ \cite{mpk2016}
(where t$_{L}$ is the kinetic energy of the light fermion species, discussed later)
while for the 6$_{Li}$-40$_{K}$ mixture T$_{c}$ $\sim$ 0.03t$_{L}$.}
\item{For T $>$ T$_{c}$, two pseudogap regimes are realized as PG-I and PG-II regime,
corresponding to regions where both and only the lighter fermion species are 
pseudogapped, respectively. Close to unitarity, the heavy species is 
pseudogapped upto T$\sim$58.5nK, while in the light species the pseudogap 
survives upto T $>$ 108nK.}
\item{In presence of population imbalance, uniform superfluid state 
(with zero momentum pairing) is 
realized only beyond a critical interaction U$_{c}$. This is in remarkable contrast 
to the balanced case where any arbitrarily small attractive interaction 
gives rise to a stable uniform superfluid.}
\item{Imbalance in population leaves it's imprint on the superfluid gap. 
In spite of an isotropic s-wave interaction finite momentum scattering gives rise 
to ``nodal'' superfluid gap.}
\end{itemize}

Rest of the paper is organized as follows. In section II we discuss about 
the model and the numerical method that has been used to study the 
imbalanced Fermi mixture. Section III discusses our results 
for population balanced and imbalanced systems with imbalance in mass. 
We further present quantitative 
estimates of the thermal scales corresponding to the 6$_{Li}$-40$_{K}$ mixture. 
We touch upon certain computational issues in the discussion section IV and 
conclude with section V. 

%%%%%%%%%%%%%%%%%%%%%%%%%%%%%%%%%%%%%%%%%%%%%%%%%%%%%%%%%%%%%%%%%%%%%%%
\begin{figure}
\begin{center}
\includegraphics[height=6cm,width=6cm,angle=0]{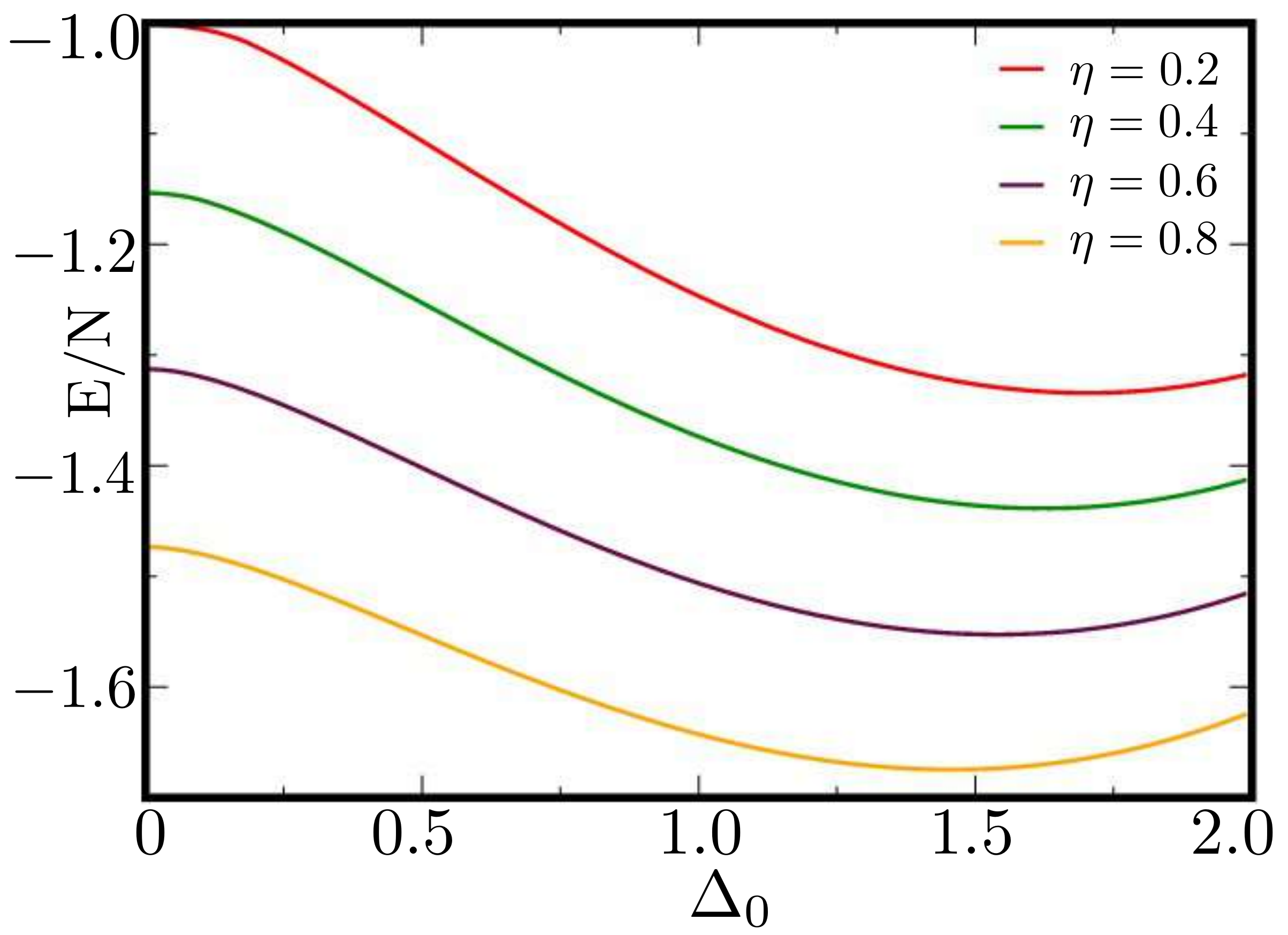}    
\end{center}
\caption{Color online: Variation of free energy density with 
pairing field amplitude at different mass imbalance ratio and zero 
population imbalance.}
\end{figure}
%%%%%%%%%%%%%%%%%%%%%%%%%%%%%%%%%%%%%%%%%%%%%%%%%%%%%%%%%%%%%%%%%%%%%%%

\section{Model, method and indicators}

\subsection{Model}

We study the attractive Hubbard model on a 
square lattice with the fermion species having 
unequal masses, additionally they are being 
subjected to an imbalance in population,
\begin{eqnarray}
 H &=& H_{0} - h\sum_{i}\sigma_{iz} - \mid U\mid
 \sum_{i}n_{i\uparrow}n_{i\downarrow}
\end{eqnarray}
where, $H_{0} = \sum_{i, j, \sigma} (t_{ij\sigma}-\mu
\delta_{ij})c_{i\sigma}^{\dagger}c_{j\sigma}$, with
$t_{ij\sigma} = -t_{\sigma}$ only for nearest neighbor
hopping and is zero otherwise. $\sigma$ correspond to the 
$\uparrow$ and $\downarrow$ spin species of fermions,
henceforth referred as $L$ and $H$, respectively;
where $H$ stands for the heavy fermion species and $L$
for the lighter one. $t_{\sigma} \propto 1/m_{\sigma}$, 
takes into account the unequal masses of the fermion 
species; $t_{L}$ serves as the energy scale in the problem in terms 
of which the various quantities are measured.
We define the mass imbalance ratio as $\eta = 
m_{L}/m_{H}$, where $m_{H}$ and $m_{L}$ 
are the effective masses of the two species. $\eta = 1$
thus correspond to the mass balanced situation. We measure the 
population imbalance in terms of an ``effective field''
$h=(1/2)(\mu_{L} - \mu_{H})$, where $\mu_{L}$
and $\mu_{H}$ correspond to the chemical potential of the 
light and heavy fermion species, respectively. The polarization 
is defined as $m=\langle n_{L}^{i} - n_{H}^{i}\rangle$, with 
$n^{i}$'s being the corresponding number density of the fermion species.
 
For the system under consideration we want to
explore the physics beyond the weak coupling, which requires
one to retain the fluctuations beyond the mean field theory.
For this we use a single channel Hubbard-Stratonovich
decomposition of the interaction in terms of an auxiliary 
complex scalar field $\Delta_{i}(\tau) = \mid \Delta_{i}(\tau)
\mid e^{i\theta_{i}(\tau)}$. A complete treatment of the 
problem requires retaining the full $(i, \tau)$ dependence 
of the $\Delta$, a target that can be achieved only through 
imaginary time QMC. 
The present technique known as the static auxiliary field
(SAF) Monte carlo \cite{evenson1970, avishai2007}
ignores the temporal fluctuations but 
retain the complete spatial fluctuations of $\Delta_{i}$. This approximation makes
the technique akin to the mean field theory at T=0, but retains the amplitude 
and phase fluctuations of $\Delta_{i}$ at finite  temperatures, 
which controls the thermal scales. In the language of Matsubara 
frequency, SAF retain fluctuations corresponding to 
$\Omega = 0$ mode only. A detail account of
our technique can be found in ref. \cite{mpk2016}.

The effective Hamiltonian is,
\begin{eqnarray}
 H_{eff} & = & H_{0}-h\sum_{i}\sigma_{iz} + \sum_{i}(\Delta_{i}
 c_{i\uparrow}^{\dagger}c_{i\downarrow}^{\dagger} + h.c) + H_{cl} 
\end{eqnarray}
where, $H_{cl} = \sum_{i}\frac{\mid \Delta_{i}\mid^2}{U}$ is the 
stiffness cost associated with the now ``classical'' auxiliary field.
The pairing field configurations in turn are controlled by the 
Boltzmann weight,

\begin{eqnarray}
 P\{\Delta_{i}\} \propto Tr_{c, c^{\dagger}}e^{-\beta H_{eff}}
\end{eqnarray}

This is related to the free energy of the fermions in the 
configuration $\{\Delta_{i}\}$. For large and random $\{\Delta_{i}\}$
the trace has to be computed numerically. For this we generate 
equilibrium $\{\Delta_{i}\}$ configurations by Monte Carlo 
technique, diagonalizing the fermion Hamiltonian $H_{eff}$ for 
each attempted update.

\subsection{Numerical method}

Even though MFT is frequently used to study 
the imbalance Fermi superfluids, it is essential to retain the 
crucial thermal fluctuations as one moves beyond the 
weak coupling regime . The issue has been 
widely discussed in the context of BCS-BEC crossover in 
balanced Fermi systems \cite{nozieres1985, tamaki2008, dupuis2004,
kopec2002, scalleter1989, randheria1995, allen1999, paiva2d, keller2001, capone2002,
toschi2005, toschi_other2005, garg2005}.
For analyzing the ground state and finite temperature behavior
of the mass imbalanced system we have employed variational minimization 
and a Monte Carlo simulated annealing, respectively.

\subsubsection{Simulated annealing by Monte Carlo}

The SAF scheme can access significantly larger system sizes 
($\sim 40 \times 40$) as compared to what can be accessed 
through QMC. In order to make the 
study numerically less expensive the Monte Carlo is 
being implemented through a cluster approximation \cite{skumar, mpk2016},
wherein instead of diagonalizing the entire lattice of dimension 
$L \times L$ for each attempted update we diagonalize a cluster 
of size $L_{c} \times L_{c}$ surrounding the update site. 
For most of the results presented in this paper we have used 
a lattice of size $L = 24$, with the cluster size being $L_{c} = 6$,
for typically 4000 Monte Carlo steps.

\subsubsection{Variational minimization scheme}

At zero temperature the ground state of the system is 
determined by minimizing the energy over the static 
configurations of the pairing field $\Delta_{i}$. 
The procedure is carried out over the (U, h, $\eta$)
space, for the pairing field amplitude being defined as 
$\mid \Delta_{i} \mid \propto \Delta_{0}\cos({\bf q . r_{i}})$,
which takes into account modulations in the pairing field amplitude;
here , $\Delta_{0}$ is assumed to be real and positive. ${\bf q}$ is the 
modulation wave vector and for the balanced (uniform)
superfluid phase ${\bf q} = 0$.
We have also verified the situation 
with modulations in the pairing field phases but have found it to be 
energetically unfavorable over the parameter regime under consideration.
For the regime of interest $h_{c1} < h < h_{c2}$ (where $h_{c1}$ and $h_{c2}$
are critical population imbalances, discussed later) the modulated superfluid 
state is of Larkin-Ovchinnikov (LO) type. 

\subsection{Parameter regime and indicators}

For the results presented in this paper the interaction 
(U=4t$_{L}$) is set to be close to unitarity, unless specified otherwise.
The implementation 
of a real space simulation technique leads to 
restriction on the system sizes that can be accessed. 
Smaller interactions (U $\leq 2t_{L}$) requires larger 
system sizes since the T=0 coherence length $\xi_{0}$ 
becomes large. Setting U = 4t$_{L}$ we have explored the mass 
imbalance over the regime $\eta \sim [0:1]$ and population
imbalance $h/t_{L} \sim [0:1.50]$. The calculations 
are carried out at a fixed net chemical potential of 
(1/2)($\mu_{L}+\mu_{H}$) = -0.2t$_{L}$.
Along the selected cross sections across the parameter space 
we characterize the phases based on the following
thermodynamic and quasiparticle indicators, (i) pairing field structure 
factor (S$_{\Delta}({\bf q})$), (ii) polarization 
(m = $\langle n_{L}^{i} - n_{H}^{i}\rangle $),
(iii) pair correlation ($\Gamma({\bf q})$)
(iv) momentum resolved spectral function $A_{\sigma}({\bf k}, \omega)$ 
(where $\sigma$ = L, H), 
(v) low energy spectral weight $A({\bf k}, 0)$ distribution at the Fermi level,
(vi) species resolved occupation number (n$_{\sigma}({\bf k})$) 
and
(vii) species resolved fermionic density of states (DOS) ($N_{\sigma}(\omega)$). We define these indicators below,

\begin{eqnarray}
 S_{\Delta}({\bf q}) & = & \frac{1}{N^{2}}\sum_{i,j} \nonumber
 \langle \Delta_{i}\Delta_{j}^{*}\rangle e^{i{\bf q}.({\bf r}_{i}-{\bf r}_{j})} \nonumber 
\cr \Gamma({\bf q}) & = & \sum_{ij}\Gamma_{ij}e^{i{\bf q}.({\bf r}_{i}-{\bf r}_{j})}, 
where,\Gamma_{ij}=\langle c_{iH}^{\dagger}c_{iL}^{\dagger}\rangle
\langle c_{jL}c_{jH} \rangle
\cr A_{\sigma}({\bf k}, \omega) & = & -(1/\pi)Im G_{\sigma}({\bf k}, \omega)
\cr A_{\sigma}({\bf k}, 0) & = & -(1/\pi)Im G({\bf k}, \omega \rightarrow 0)
 \cr N_{L}(\omega) & = & \langle (1/N)\sum_{i,n}\mid u_{n}^{i}\mid^{2}
\delta(\omega - E_{n})\rangle \nonumber 
\cr N_{H}(\omega) & = & \langle (1/N)\sum_{i,n}\mid v_{n}^{i}\mid^{2}
\delta(\omega + E_{n})\rangle \nonumber
\end{eqnarray}
Here, $G_{\sigma}({\bf k}, \omega) = lim_{\delta \rightarrow 0}G_{\sigma}({\bf k}, i\omega_{n})
\mid_{i\omega_{n}\rightarrow \omega+i\delta}$ where $G_{\sigma}({\bf k}, i\omega_{n})$
is the imaginary frequency transform of $\langle c_{\bf k \sigma}(\tau)c_{\bf k \sigma}^{\dagger}(0) \rangle$.
$u_{n}^{i}$ and $v_{n}^{i}$ are the BdG eigenvectors corresponding to the eigen values $E_{n}$
for the configurations under consideration. $N = L^{2}$ is the number of lattice sites. 

\section{Results}

In this section we discuss the results obtained through our numerical simulations. We categorize our observations
in two groups viz. (i) mass imbalanced Fermi-Fermi mixture with balanced population and
(ii) mass imbalanced Fermi-Fermi mixture with population imbalance. Within each category
we discuss the ground state and finite temperature behavior separately.

%%%%%%%%%%%%%%%%%%%%%%%%%%%%%%%%%%%%%%%%%%%%%%%%%%%%%%%%%%%%
\begin{figure}
\begin{center}
\includegraphics[height=5cm,width=8cm]{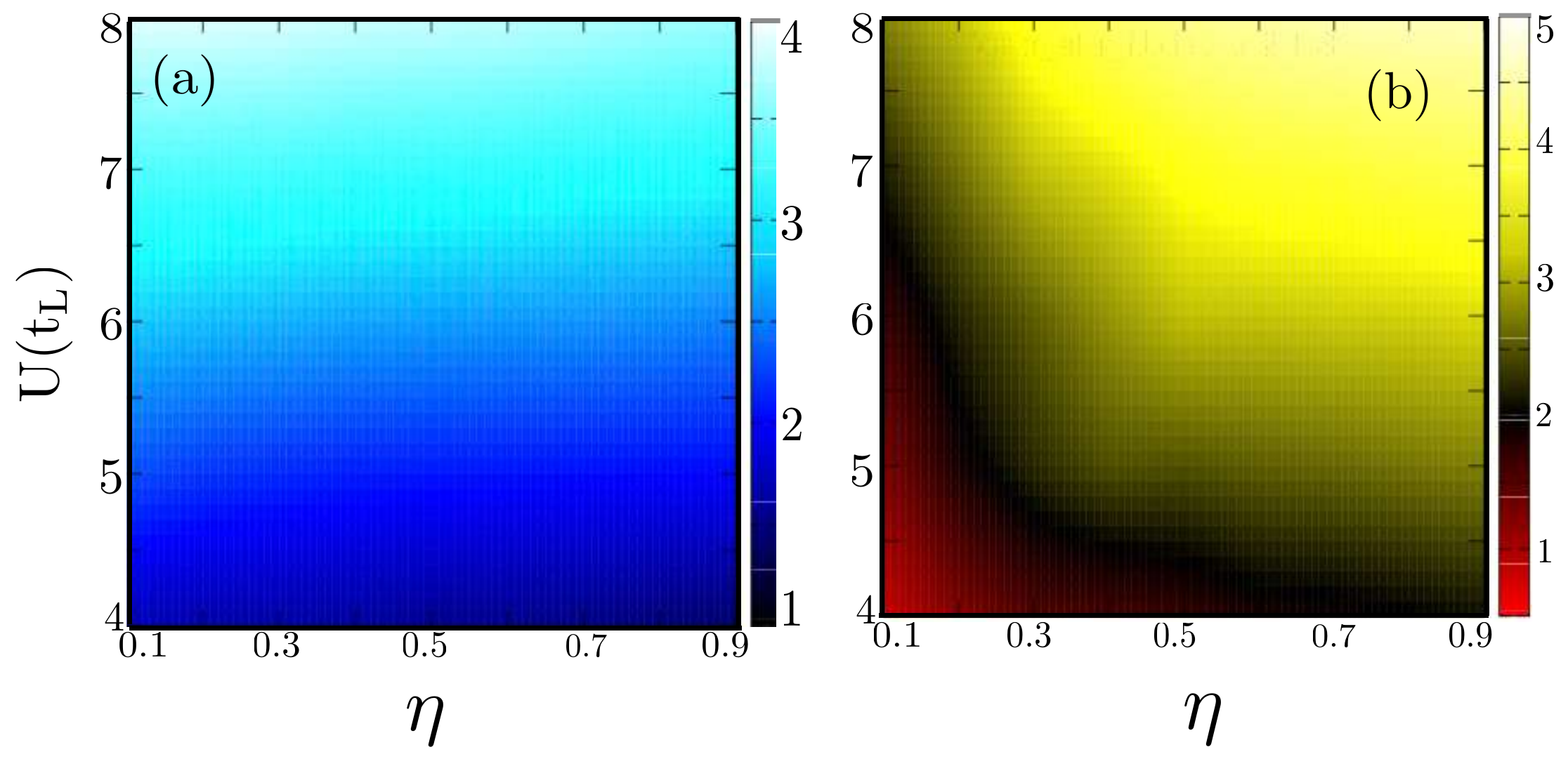}
\end{center}
\caption{Color online: Ground state (a) pairing field amplitude and 
(b) superfluid
gap in the $\eta$-U plane for the population balanced system. 
Strong interaction and small mass imbalance ($\eta \rightarrow 1$)
leads to a larger pairing field amplitude and gap at the Fermi 
level.}
\end{figure}
%%%%%%%%%%%%%%%%%%%%%%%%%%%%%%%%%%%%%%%%%%%%%%%%%%%%%%%%%%%%

\subsection{Population balanced Fermi-Fermi mixture}

Population balanced Fermi gas correspond to the situation when the fermionic
species are being subjected to equal chemical potential. In 
the context of ultracold atomic gases such a system is realized by loading 
equal population of two fermion species in the optical lattice. 

\subsubsection{Ground state}

We begin the discussion of our results in this section with the effect of mass imbalance on the ground state of the system.
With $h$ being set to zero we use the mass
imbalance ratio $\eta$ as the tuning parameter to investigate the various properties. 
We map out the ground state phase diagram based on the variational mean field calculation (discussed in the previous section) and in Fig.1 show the dependence of free 
energy on the pairing field amplitude $\Delta_{0}$ and mass 
imbalance ratio $\eta$ for the selected interaction strength U=4t$_{L}$.
The ground state energy shows a single minima and correspond to an uniform superfluid
state with the pairing field amplitude ($\Delta_{0}$) being almost independent  of the choice of the mass imbalance
ratio. Both the pairing field 
amplitude as well as the superfluid gap at the Fermi level increases
monotonically with  U.  We show these behavior in the $\eta$-U plane in
Fig.2a and 2b, respectively.

The effect of mass imbalance on the quasiparticle dispersion spectra is probed next and we 
show the corresponding spectral function A(${\bf k}, \omega$) at
different mass imbalance ratio in Fig.3. The ${\bf k}$-summed quantity of the spectral function correspond to
the electronic density of states (DOS) and we show the species resolved variant of the same (N$_{\sigma}(\omega)$) as the last two panels of Fig.3.
For the computation of both the spectral function and the DOS we have used the Green’s function formalism 
which gives access to large system sizes and thus makes the van Hove singularities prominent. Details of the Greens function formalism are discussed in section IV. 

We observe that at large imbalance in mass there are essentially four branches in the dispersion
spectra. Both above and below the Fermi level the branches intersect each other at ${\bf k}$ = $\pi$/2, 
which gives rise to additional singularities in the form of subgap and supergap states in the DOS. 
However, there is no spectral weight at the Fermi level which ensure that the underlying ordered state
is gapped across the
range of $\eta$. In a later section we would find that this behavior is remarkably
altered once a population imbalance in introduced in the system.
With decreasing mass imbalance the branches merge together, leading to the well known
two-branched BCS spectra as $\eta \rightarrow$ 1 \cite{gaebler_expt}.

In agreement with the behavior of the spectral function we observe that for the system at or 
close to the mass balanced situation, the species resolved DOS exhibits prominent hard gap at the 
Fermi level separated by the characteristic BCS-like coherence peaks. Increasing
imbalance gives rise to subgap and supergap states.

%%%%%%%%%%%%%%%%%%%%%%%%%%%%%%%%%%%%%%%%%%%%%%%%%%%%%%%%%%%%%
\begin{figure*}
\begin{center}
\includegraphics[height=5.5cm,width=16.5cm,angle=0]{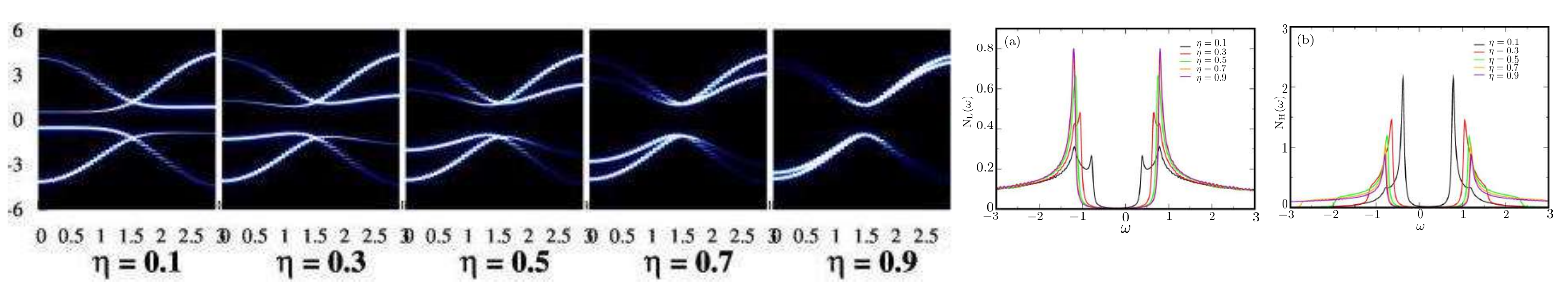}
\end{center}
\caption{Color online: Ground state dispersion spectra at different mass imbalance 
ratio $\eta$. The last two panels show the $\eta$ dependence of the 
density of states (DOS) for the (a) light and (b) heavy fermion species.
Note the sub and super gap features in the DOS at large mass imbalance ratio.}
\end{figure*}
%%%%%%%%%%%%%%%%%%%%%%%%%%%%%%%%%%%%%%%%%%%%%%%%%%%%%%%%%%%%%

\subsubsection{Finite temperature}

Thermal evolution of this system is probed in terms of two thermodynamic quantities viz. (i) the  
pairing field structure factor (S(${\bf q}$)) and (ii) the pair correlation 
($\Gamma$(${\bf q}$)). While both these quantities essentially give similar 
information, $\Gamma$(${\b q}$) is a more fundamental quantity since it 
directly probes the fermionic correlations rather than the correlation 
between the auxiliary fields. Spatial maps of pairing field structure factor as well as pair correlation (not shown here) exhibit an uniform 
superfluid (BCS-like) low temperature state with a finite peak 
at ${\bf q} = 0$.  
Based on these two quantities we map out 
the mass imbalance-temperature ($\eta$-T) phase diagram at U=4t$_{L}$
and show it in Fig.4a. 
Note that in a two-dimensional system such as the present one, 
thermal transitions are possible only through a Berezinskii-Kosterlitz-Thouless
 (BKT) transition. The T$_{c}$ scales discussed here correspond 
to the BKT transition temperature.

There are two thermal scales in this phase diagram viz. T$_{c}$
and T$^{*}$ which demarcates the phases as superfluid, pseudogap and normal. While T$_{c}$ corresponds 
to the temperature beyond which the phase coherence in the pairing field is lost, T$^{*}$ marks the loss of short range pair correlations leading to the 
disappearance of superfluidity. 

Thermal fluctuations are progressively detrimental with increasing 
mass imbalance in the system, leading to suppression in T$_{c}$ as shown in Fig.4a. The observation
 suggests 
that even though at the ground state there is a large pairing field amplitude the state is fragile
towards thermal fluctuations and rapidly looses phase coherence. The thermal scales T$_{c}$ and T$^{*}$
are determined from the pairing field structure factor peaks 
(S$_{\Delta}$(${\bf q}$)) shown in Fig.4b.
Away from the weak coupling regime the short range pair correlations survive upto temperatures T$^{*}$ $>>$ T$_{c}$.
In the limit of $\eta \rightarrow$ 1 we find T$^{*}$ $\approx$ 2T$_{c}$.
The pseudogap regime is 
determined based on the temperature dependence of the structure factor and pair correlation peak at ${\bf q} = 0$ and T$^{*}$
correspond to the temperature at which there is no distinguishable peak in the S(${\bf q}$) and $\Gamma$(${\bf q}$) at ${\bf q} = 0$.
Fig.4a is one of the important results of this work. It shows how the consideration of 
phase fluctuations in the numerical framework is important to capture the
true thermal scales of the
mass imbalanced superfluid system. A finite temperature MFT
can track only T$^{*}$ and thus lead to significant overestimation of
T$_{c}$.
However, even though the thermodynamic measures S(${\bf q}$) and 
$\Gamma$(${\bf q}$) are sufficient to quantify the existence of phase coherence in the system, 
they lack the information about the quasiparticle behavior of the fermionic species. Analyzing the 
fermionic 
properties such as single particle DOS, spectral function etc. (discussed later) 
shows us how the information about the quasiparticle behavior significantly alters the
 phase diagram in Fig.4a. 

In Fig.4c we show the BCS-BEC crossover at a selected mass imbalance ratio of $\eta$ = 0.15, 
corresponding to the experimentally realized Fermi-Fermi mixture of 6$_{Li}$-40$_{K}$ \cite{taglieber2008, naik2010, wille2008, costa2010, naik2010}.
 Further, 
we compare our result with the one obtained for a mass balanced system, so as to demonstrate the 
suppression of T$_{c}$ by imbalance in mass.  

Across the BCS-BEC crossover the behavior of the T$_{c}$ scale is
governed by different mechanisms at different coupling regimes.   
In the weak coupling regime the thermal 
scale is determined by the vanishing of the pairing amplitude ($\langle \langle c_{iL}^{\dagger} 
c_{iH}^{\dagger}\rangle \rangle$) as $k_{B}T \sim te^{-t/U}$. At strong interactions where 
the system comprises of molecular pairs the thermal scale is dictated by the phase correlation of the 
local order parameter and behaves as $k_{B}T \sim f(n)t^{2}/U$, where $f(n)$ is
a function of number density.

In Fig.4d we show the composite thermodynamic phase diagram in the $\eta$-U-T space. A large imbalance in mass suppresses the
T$_{c}$ scale irrespective of the choice of U. For eg., at U=4t$_{L}$, T$_{c}$ $\sim$ 0.16t$_{L}$ at $\eta$ = 0.9, and
progressively reduces to T$_{c}$ $\sim$ 0.13t$_{L}$ at $\eta$ = 0.5 and to T$_{c}$ $\sim$ 0.04t$_{L}$ at $\eta$ = 0.1,
respectively. The BCS-BEC crossover remains roughly unaffected (except for this suppression)
 by the change in the mass imbalance ratio, with
U$_{c}$ $\sim$ 5t$_{L}$ corresponding to unitarity with maximum T$_{c}$ \cite{paiva2d}. 
We would come back to the concept of unitarity in a lattice fermion model in the discussion section of 
this paper. 

The quasiparticle behavior is discussed next and in Fig.5 we show the effect of mass imbalance on species resolved DOS.
The effect of mass imbalance on the DOS can be observed through multiple features. A quick look at the
magnitude of the coherence peaks of the DOS in Fig.5 shows that the heavy fermion species has it’s coherence
peaks significantly larger in magnitude as compared to it’s lighter counterpart. In Fig.5 we 
have shown the species resolved DOS corresponding to three different mass imbalance ratio as 
$\eta=0.2$ (panels (a) and (d)), $\eta=0.5$ (panels (b) and (e)) and $\eta=0.9$ (panels (c) 
and (f)). While the difference in magnitude of the coherence peaks between the two species is 
maximum at large mass imbalance, it progressively reduces as the system transits to the 
balanced situation and at $\eta=0.9$ they are almost equal, as expected from the mass balanced 
situation. Secondly, we observe that the two species are now being subjected to different scaled
temperatures and the heavy species experiences a higher temperature as compared to the lighter 
ones. This is because the kinetic energy contribution of the two species are now different in 
presence of imbalance in mass. {\it Thus, there are now two different pseudogap scales
in the system} and one needs species resolved probes such as rf spectroscopy to access 
them.  Finally, at larger imbalance in mass the pseudogap behavior is almost 
independent of thermal evolution and persists even at high temperatures. It is only close to 
the mass balanced situation that thermal fluctuations begin to pile up significant weight 
at the Fermi level. The observation is crucial and suggests that in case of Fermi-Fermi 
mixtures thermodynamic quantities such as pairing field structure factor (Fig.4a) 
significantly underestimates the pseudogap regime. Information about the quasiparticle 
behavior is essential in this case in order to obtain the complete picture of 
the thermal behavior of the system. In the later sections we would observe that inclusion
of population imbalance is instrumental in making such mixtures reactive towards temperature
and even with large imbalance in mass the pseudogap undergoes significant thermal evolution. 
This can be summed up as that while an imbalance in mass promotes the pseudogap 
behavior, an imbalance in population leads suppression of the pseudogap scales.        
%%%%%%%%%%%%%%%%%%%%%%%%%%%%%%%%%%%%%%%%%%%%%%%%%%%%%%%%%%
\begin{figure}
\begin{center}
\includegraphics[height=8cm,width=8.5cm]{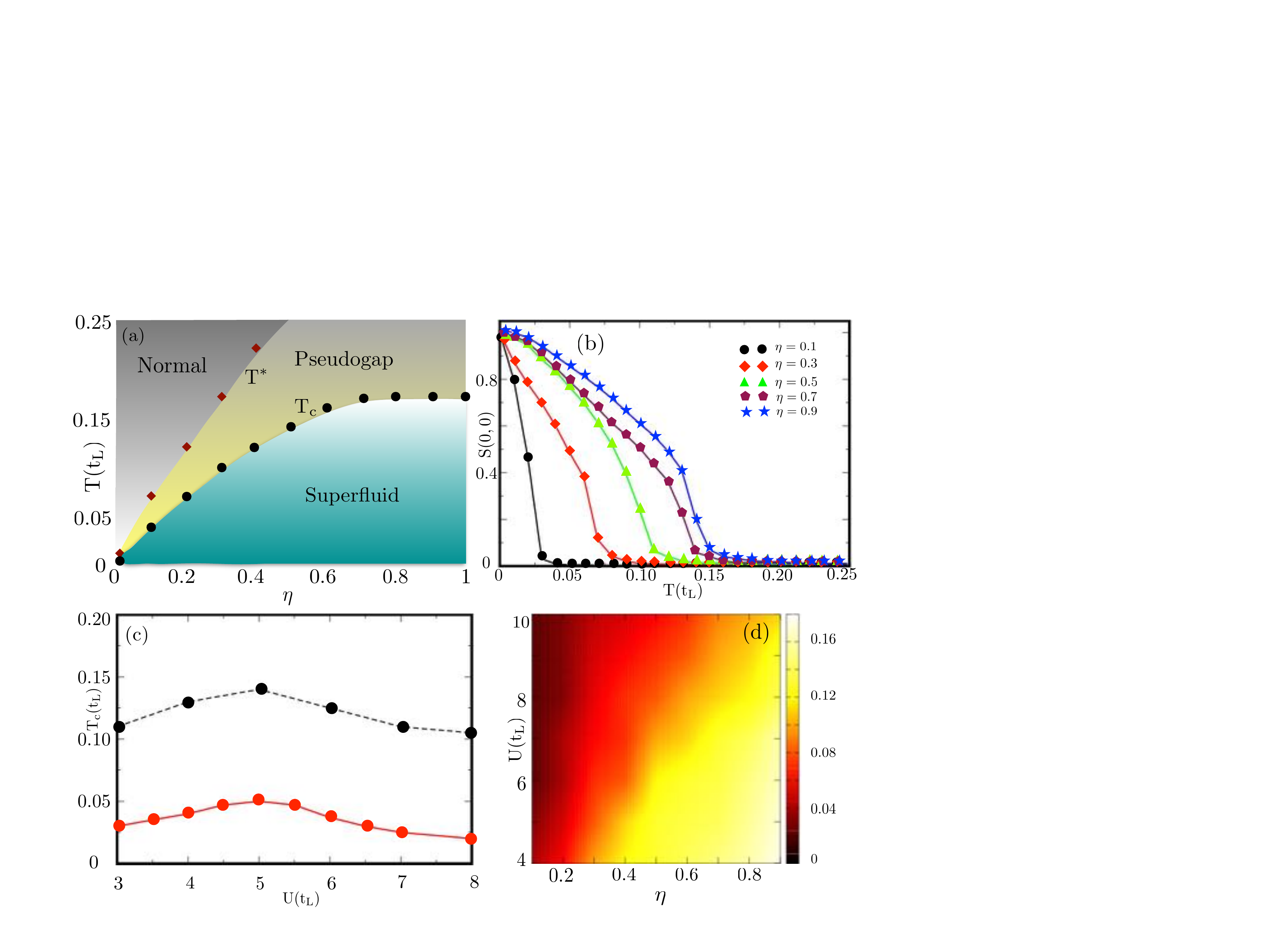}
\end{center}
\caption{Color online: (a) Effect of mass imbalance ($\eta$) on the thermal 
scales (T$_{c}$ and T$^{*}$) at U=4t$_{L}$, for a population balanced 
system. T$_{c}$ corresponds to the temperature at which the system 
looses it's phase coherence. Beyond T$^{*}$ there is no noticeable peak
 in the pairing 
field structure factor. (b) Thermal evolution of pairing field structure 
factor peak at 
different mass imbalance ratio. (c) BCS-BEC crossover at $\eta = 0.15$ (red solid),
for a population balanced Fermi-Fermi mixture. Note the suppression in T$_{c}$ 
due to mass imbalance, in comparison to the balanced case (black dotted line). 
(d) BCS-BEC crossover in the $\eta$-U plane.
The balanced situation corresponds to a large pairing field amplitude and thus a 
higher T$_{c}$.}
\end{figure}
%%%%%%%%%%%%%%%%%%%%%%%%%%%%%%%%%%%%%%%%%%%%%%%%%%%%%%%%%%

A second quasiparticle behavior of interest is the momentum resolved spectral function A(${\bf k},\omega$) and we
present the species resolved version of the same for ${\bf k}$ = $\{0, 0\}$ to $\{\pi, \pi\}$ scan
across the Brillouin zone, in Fig.6. While at the low temperature both the
species possess prominent spectral gap at the Fermi level, progressive increase in temperature smears out the gap. The thermal disordering temperature corresponding to the two species are now different, leading 
to two pseudogap scales. Since t$_{H}$ $<$ t$_{L}$, the heavy species experiences a higher ``scaled'' 
temperature and consequently undergoes faster thermal disordering. It must however be noted that
even at high temperature there is a very small but noticeable gap at the Fermi level in 
agreement with the behavior of the DOS.  
%%%%%%%%%%%%%%%%%%%%%%%%%%%%%%%%%%%%%%%%%%%%%%%%%%%%%%%%%%
\begin{figure*}
\begin{center}
\includegraphics[height=7cm,width=14cm,angle=0]{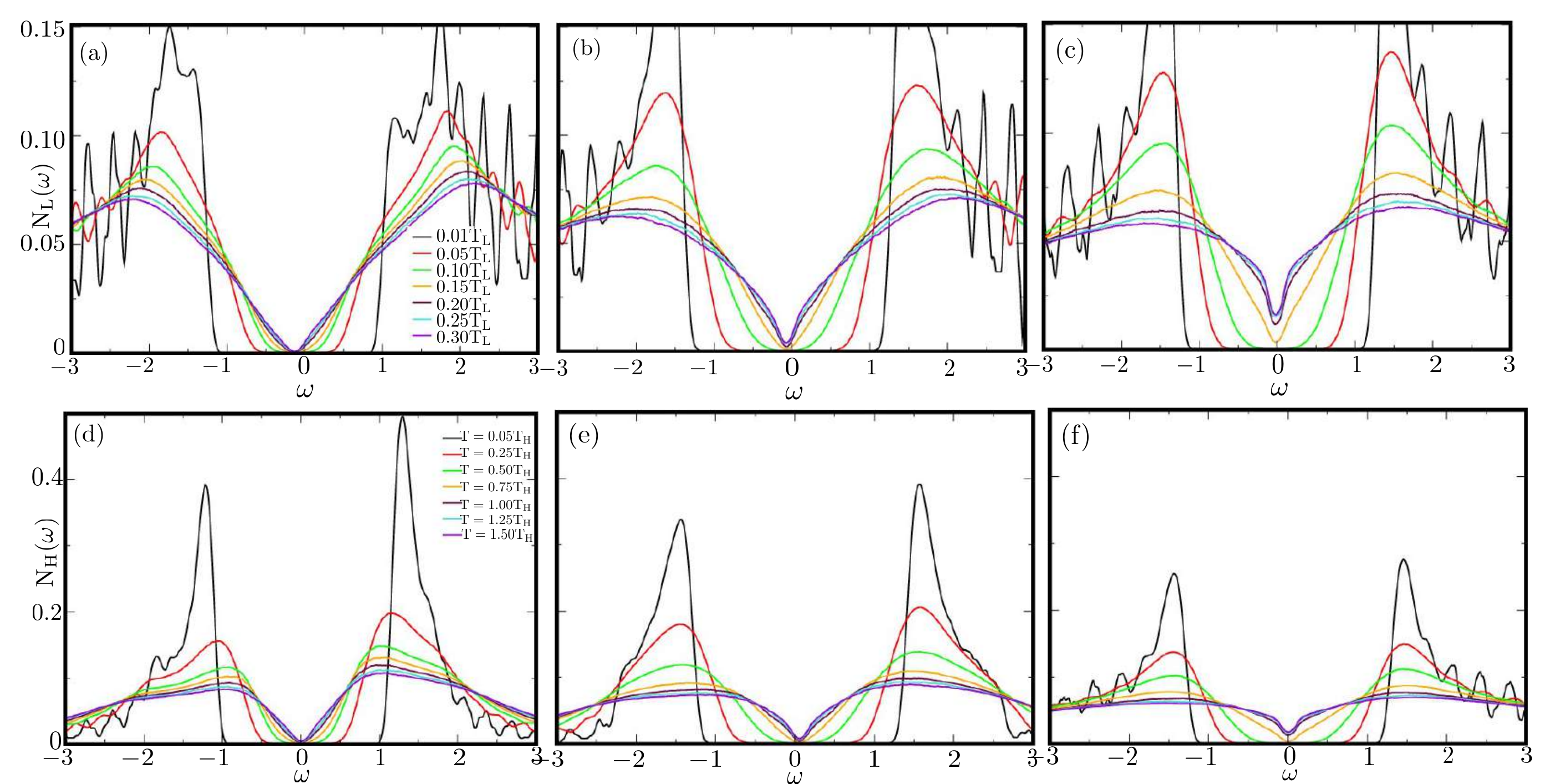}
\end{center}
\caption{Color online: Thermal evolution of DOS at the Fermi level 
for the light ((a)-(c)) and the heavy ((d)-(f)) fermion 
species for different mass imbalance ratio of $\eta=0.2$ (panel (a) and (d)), 
$\eta = 0.5$ (panels (b) and (e)) and $\eta = 0.9$ (panels (c) and (f)).
At large imbalance in mass the DOS corresponding to the heavy species has coherence 
peaks with magnitudes significantly larger than it's light counterpart. As the 
system approaches the mass balanced situation the coherence peaks of the heavy 
species reduces and for $\eta \rightarrow 1$ becomes equal to that of the
 light species.}
\end{figure*}
%%%%%%%%%%%%%%%%%%%%%%%%%%%%%%%%%%%%%%%%%%%%%%%%%%%%%%%%%%
 A species dependent momentum 
resolved photoemission spectroscopy measurement is a suitable experimental technique to observe the species dependent thermal disordering of the 
spectral functions. 
While the signature of short range pair correlation at T $>$ T$_{c}$ in both the species merely 
demonstrates the loss of pair coherence, the survival of such short range pair correlation {\it only} in the light species would be a true signature of imbalance in mass in the system. 

%%%%%%%%%%%%%%%%%%%%%%%%%%%%%%%%%%%%%%%%%%%%%%%%%%%%%%%%%%
\begin{figure*}
\begin{center}
\includegraphics[height=8cm,width=15cm,angle=0]{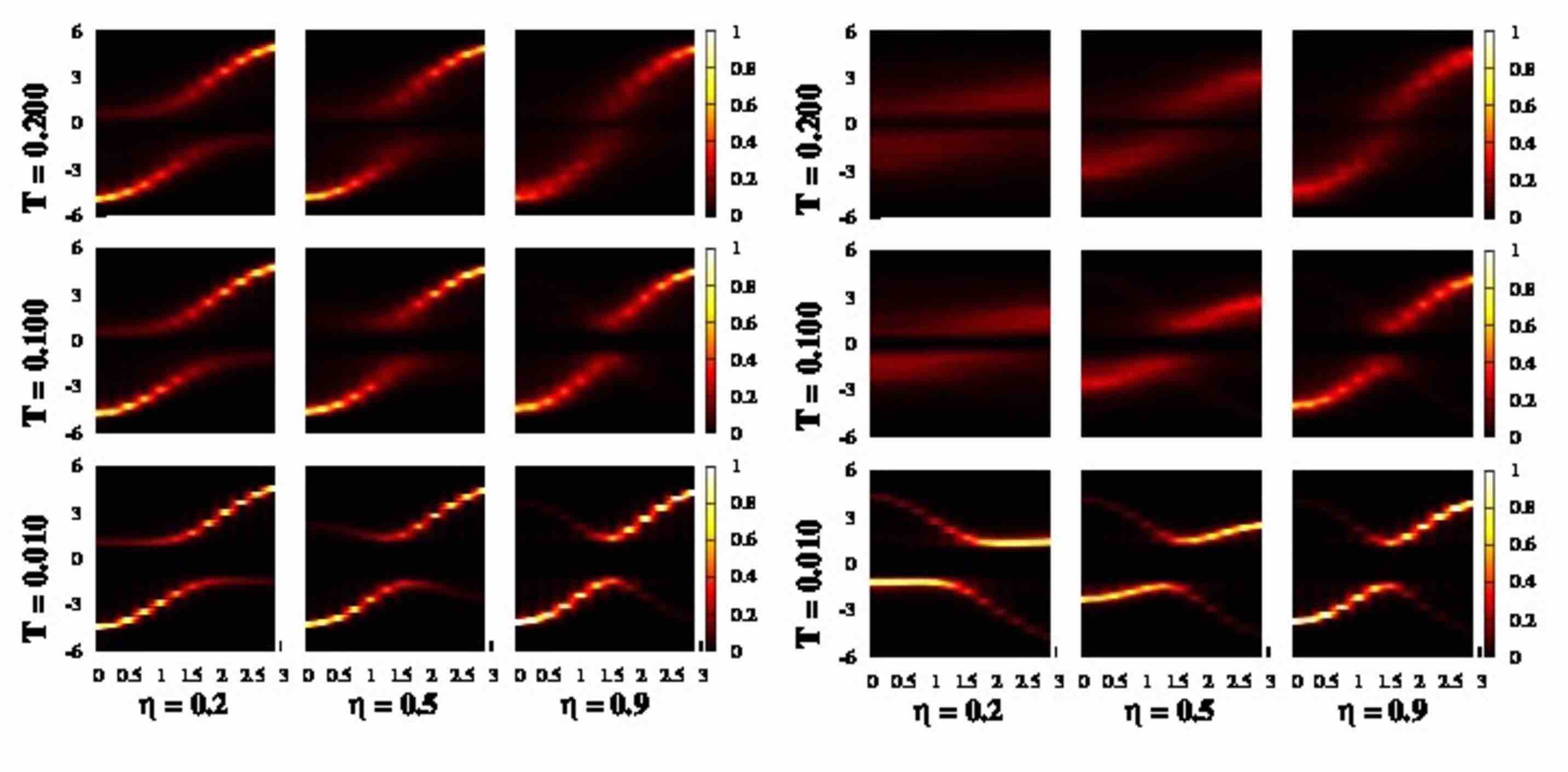}
\end{center}
\caption{Color online: Thermal evolution of species resolved 
(light (left) and heavy (right)) spectral function A(${\bf k}, \omega$) 
at selected $\eta$. All temperatures are measured in terms of t$_{L}$, 
where t$_{L}$=t$_{H}$/$\eta$. Since the heavy species experiences a 
higher ``scaled'' temperature (see text) it undergoes faster thermal disordering.}
\end{figure*}
%%%%%%%%%%%%%%%%%%%%%%%%%%%%%%%%%%%%%%%%%%%%%%%%%%%%%%%%%%

\subsection{Population imbalanced Fermi-Fermi mixture}

We now introduce the next level of complexity to the system by adding on 
an imbalance in population along with the existing mass imbalance. 
Before proceeding further we quickly summarize how the imbalance in 
population in introduced in our model.
An imbalance in population can be created through a mismatch in the size 
of Fermi surface corresponding to the two fermionic species. This in turn 
can be achieved in two ways, viz. (i) by creating difference in the chemical 
potential or (ii) by creating difference in the number density of the two 
fermionic species.  

%%%%%%%%%%%%%%%%%%%%%%%%%%%%%%%%%%%%%%%%%%%%%%%%%%%%%%%%%%%%%%%%%%%%%%%%%%%%%%%%%%                        
\begin{figure*}
\begin{center}
\includegraphics[height=6cm,width=15cm,angle=0]{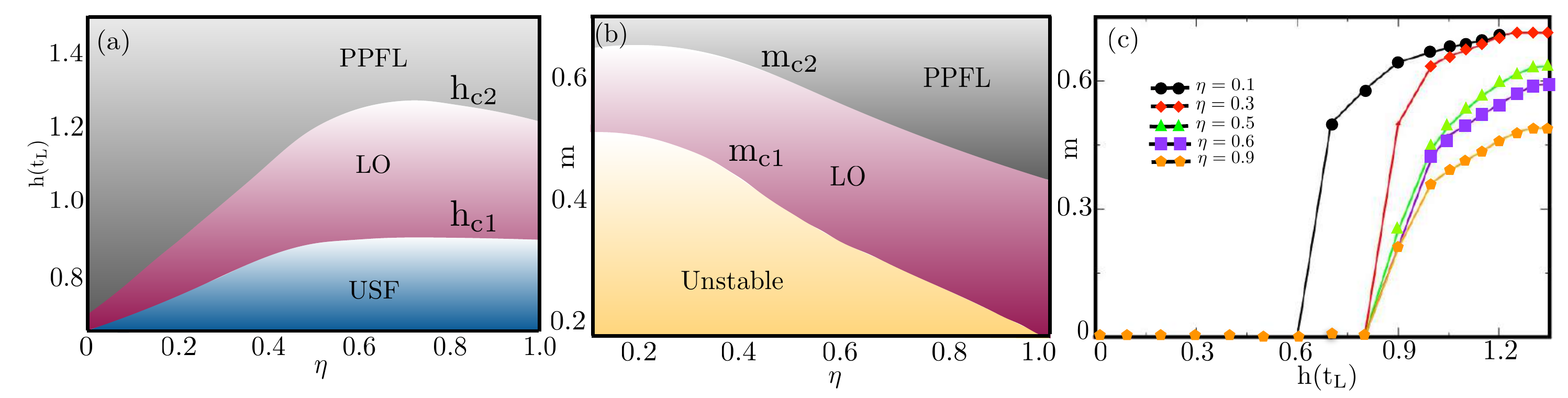}
\end{center}
\caption{Color online: Ground state phase diagram for the 
mass imbalanced system as a function of varying population imbalance 
at U=4t$_{L}$, in the (a) $\eta$-h and
(b) $\eta$-m plane. Upto a critical effective field (h$_{c1}$) say, the system
is an unpolarized superfluid (USF). A small $\eta$ gives rise to a
small h$_{c1}$ and beyond $\eta \sim 0.5$, h$_{c1}$
becomes independent of the choice of $\eta$. For the range of field
$h_{c1} < h < h_{c2}$ a modulated superfluid (LO) phase is realized
for any choice of $\eta$. h$_{c2}$ is only weakly dependent on the
choice of $\eta$ beyond $\eta \sim 0.5$. For $h > h_{c2}$ the ground
state of the system is a partially polarized Fermi liquid (PPFL).
In the $\eta$-m plane the entire USF regime collapses on to the m=0
axis. The ground state is phase separated (unstable) in this regime
 and undergoes
first order transition to the LO state beyond a critical polarization 
m$_{c1}$, say.
A second order transition from the LO to the PPFL state takes place at
m$_{c2}$. Panel (c) shows how the polarization (m) varies with the effective
field (h) at different mass imbalance ratio $\eta$. At the critical effective
field h$_{c1}$ there is a discontinuous transition between the zero and
finite polarization state. The transition is only weakly first order
at small mass imbalance (as $\eta \rightarrow 1$), while a sharp
first order transition is realized in presence of large imbalance in
mass.}
\end{figure*}
%%%%%%%%%%%%%%%%%%%%%%%%%%%%%%%%%%%%%%%%%%%%%%%%%%%%%%%%%%%%%%%%%%%%%%%%%%%%%%%%%% 
As already mentioned the imbalance in chemical potential is quantified in 
terms of an ``effective field'' $h$, while a finite polarization $m$
is a measure of imbalance in the number densities. The quantification in terms of finite polarization is more suitable in the context of cold atomic 
experiments where the population of individual fermionic species to be loaded in the optical 
lattice can be controlled. For the solid state systems the imbalance in population is 
achieved by applying a Zeeman magnetic field which leads to a chemical potential mismatch. In our work we take this second route and subject the system to 
an effective Zeeman field ($h \neq 0$), i. e. we  
control the chemical potential to which the individual species are being subjected to, rather 
than controlling the number density of each species. The presence of
population imbalance is expected to suppress the thermal scales at any mass imbalance. In a mass balanced 
system  at a 
sufficiently large imbalance in the population an uniform superfluid state 
can not be realized. It was found that rather than transiting to a 
polarized Fermi liquid phase the system undergoes transition to a modulated 
superfluid state with finite momentum (${\bf q} \neq 0$) pairing, known 
as the Fulde-Ferrell-Larkin-Ovchinnikov (FFLO) state \cite{ff,lo,mpk2016, mpk2016_epjd}.
At weak imbalance in population the system undergoes thermal
evolution to a breached pair (BP) state comprising of coexisting superfluidity and finite polarization, a phase which does not have a ground state counterpart
\cite{mpk2016}. In the next few sections we discuss the effects of the interplay between population and mass imbalances in the system under 
consideration.

\subsubsection{Ground state}

As in the population balanced case we begin the discussion with the ground state behavior of the system
with imbalance in population and mass. The ground state phase diagram in terms of the effective field
$h$ and mass imbalance ratio $\eta$ is shown in Fig.7a. The thermodynamic phases are classified
as unpolarized superfluid (USF) (with $\Delta = 0$ and m=0), modulated superfluid (LO) 
(with $\Delta \neq 0$ and m $\neq$ 0) and partially polarized Fermi liquid (PPFL) (with 
$\Delta = 0$ and m $\neq$ 0). 
There are two critical fields in this phase diagram 
viz. h$_{c1}$ which correspond to a first order transition from the USF to the LO state
and h$_{c2}$ at which the LO state undergoes a second order transition to the PPFL phase.
%%%%%%%%%%%%%%%%%%%%%%%%%%%%%%%%%%%%%%%%%%%%%%%%%%%%%%%%%%%%%%%%%%%%%%%%%%%%%%%%%%

From the perspective of the cold atom experiments, we show the ground state phase diagram  in Fig.7b 
in the polarization-mass imbalance (m-$\eta$) plane. Note
that when depicted in terms of polarization the entire USF regime corresponding to m=0 collapses 
to the x-axis.
The first order transition from USF to LO is marked by discontinuity in the polarization, which is shown as the
unstable (phase separated) region in the Fig.7b. We observe that the discontinuity in the polarization gets
progressively enhanced with increasing imbalance in mass. Irrespective of the system size under
consideration a larger mass imbalance favors a first order transition between USF and LO phases. A
quantitative measure of this discontinuity in polarization can be seen in Fig.7c, where we relate the effective
field $h$ to the corresponding polarization $m$. 

%%%%%%%%%%%%%%%%%%%%%%%%%%%%%%%%%%%%%%%%%%%%%%%%%%%%%%%%%%%%%%%%%%%%%%%%%%%%%%
\begin{figure*}
\begin{center}    
\includegraphics[height=6cm,width=15cm,angle=0]{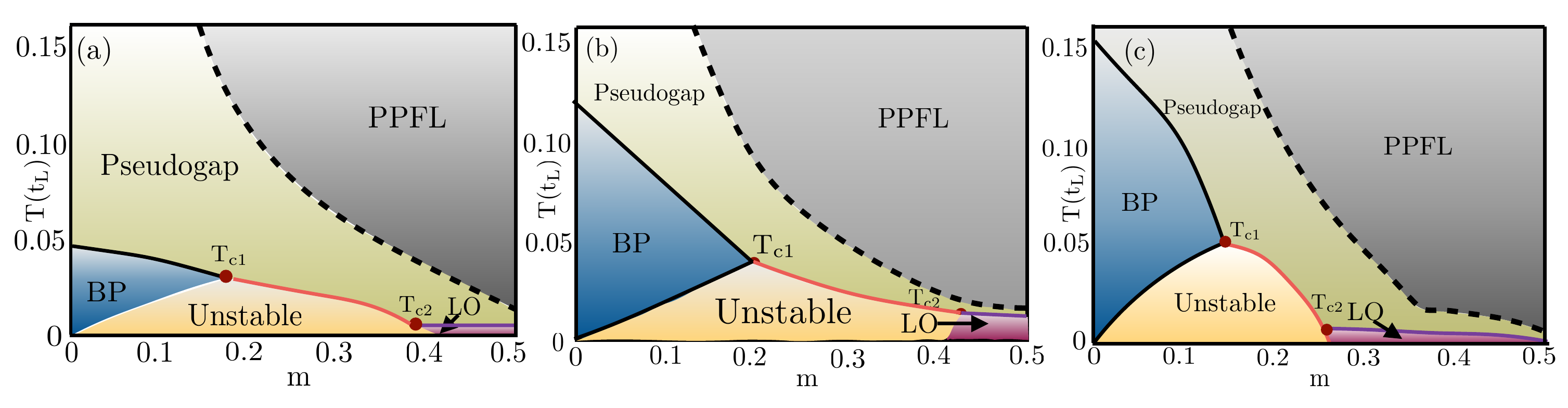}
\end{center}
\caption{Color online: Polarization-temperature (m-T) phase diagram at 
selected mass imbalance ratio of $\eta=$ (a) 0.2, (b) 0.4 and (c) 0.6.
The dashed line correspond to T$^{*}$ in each panel.
There are three broad thermodynamic phases in each case as the (i) breached
pair (BP), (ii) modulated superfluid (LO) and (c) partially polarized
Fermi liquid 
(PPFL). In the regime of weak polarization the system undergoes 
second order thermal transition (shown by black solid line) from BP to the 
pseudogap phase. At intermediate polarization the low temperature
phase separated (unstable) state undergoes a first order thermal
transition (shown by the red solid line). The large polarization regime is 
LO phase which undergoes a second order thermal transition. 
The first order transition regime is demarcated by a tricritical  
T$_{c1}$ and a Lifsitz point T$_{c2}$. A large mass imbalance in the 
system leads to significant suppression in the T$_{c}$ and thus the phase 
coherent superfluid state but gives rise to a wider pseudogap regime.}
\end{figure*}  
%%%%%%%%%%%%%%%%%%%%%%%%%%%%%%%%%%%%%%%%%%%%%%%%%%%%%%%%%%%%%%%%%%%%%%%%%%%%%%
Though not shown in the figure the LO regime is segregated into several finer regimes corresponding to the different
modulation wave vectors arising at different strength of population imbalance. The optimized  wave vector
is dictated by the lattice size, interaction strength, as well as the mass imbalance ratio.
In the variational scheme that has been used to map out the ground state phase diagram we have 
carried out the optimization of energy for different trial solutions corresponding to 
(i) uniaxial modulation, (ii) diagonal modulation and (iii) two dimensional modulations. 
For the parameter regime under consideration the uniaxially modulated LO state has been 
found to be the suitable configuration. In principle modulations with multiple wave vector 
makes up a possible candidate for the LO state, however, for the sake of numerical simplicity 
we have not allowed for such solutions in our variational scheme.

The quasiparticle spectra in the LO phase (not shown here) is pseudogapped even at the ground state.
The pseudogap behavior in this case is however a band structure effect arising out of the underlying 
modulated state.
The corresponding dispersion spectra of this phase deviates significantly from the BCS-like
behavior and is characterized by multiple dispersion branches \cite{mpk2016}. 
The multi branched dispersion spectra arises because the electrons 
now undergo finite momentum scattering  unlike the homogeneous BCS state.
The additional van Hove singularities arise from the ${\bf k}$ regions where the condition 
$\partial E_{\alpha}/\partial {\bf k}$=0 is satisfied by the dispersion spectra, where $\alpha$ correspond 
to the dispersion branches of the LO spectra \cite{mpk2016}.
While the choice of the mass imbalance ratio $\eta$ determines the optimized pairing momenta ${\bf Q}$
of the LO superfluid for a particular choice of population imbalance, the coarse features of the 
spectra (i. e. multiple branches and multiple van Hove singularities) remain unaltered by the 
choice of $\eta$.    

\subsubsection{Finite temperature}

The thermal evolution of the system is discussed in terms of m-T phase diagrams shown in Fig.8 for different choices of mass imbalance ratio $\eta$. The thermodynamic phases are determined based on the thermal
evolution of pairing field structure factor S(${\bf q}$) and polarization m(T). The broad thermodynamic
phases remain the same irrespective of the choice of $\eta$,  and with increasing imbalance in population
the system transits through a breached pair (BP), unstable, LO and PPFL phases in each case. Also
irrespective of the choice of $\eta$ there is a tricritical point T$_{c1}$ and a Lifsitz point T$_{c2}$
in the phase diagram. While T$_{c1}$ corresponds to the point where the order of transition changes 
from second to first within the BP phase, T$_{c2}$ marks the transition from the BP to the LO
phase. Unlike the continuum case \cite{stoof2010}
the two transitions are well separated
in a lattice model and the separation increases with increasing mass imbalance. 
Presence of mass imbalance significantly alters the regime of stability of the different 
thermodynamic phases. A larger imbalance in mass
leads to stronger suppression in T$_{c}$ and thus a progressively smaller BP regime. The pseudogap regime on the other hand increases monotonically with the imbalance in mass,  for example, at h=0 the ratio T$^{*}$/T$_{c}$ $\sim$ 3.33 at $\eta$ = 0.2 and reduces to 
$\sim$ 2.14 and $\sim$ 1.25 at $\eta$ = 0.4 and 0.6, respectively. 

As discussed in case of the population balanced superfluid 
the species resolved DOS continues to have different thermal disordering scales 
in the BP regime as well, owing to the different kinetic energy scales 
of the two species. The underlying superfluid state in this regime is gapped 
and undergoes thermal evolution to pseudogapped phase with increasing 
temperature. In Fig.9 we show the thermal phase diagram of the BP phase in the $\eta$-T plane 
for a particular choice of the population imbalance h=0.6t$_{L}$. 
Apart from the T$_{c}$ there are additional thermal scales in this phase diagram based on the quasiparticle behavior. We discuss them below.     

For a population imbalanced system the species resolved DOS at the shifted Fermi level 
($\omega$ = $\pm$h) shows a non monotonic behavior \cite{mpk2016}. Increasing temperature leads 
to progressive filling up of the gap upto a temperature T$_{max}$. For T $>$ T$_{max}$ a non 
monotonic thermal evolution sets in and there is now depletion of spectral weight at the Fermi 
surface with increasing temperature. 
In presence of strong interaction the pseudogap continues to survive upto high temperatures but the scale
T$_{max}$ rapidly collapses with increasing population imbalance \cite{mpk2016}.
Since T$_{max}$ survives upto temperatures
significantly higher than the T$_{c}$ it is more likely to be accessible to the experimental probes.

In presence of mass imbalance the scale T$_{max}$ is now dependent on the fermion species as T$_{max}^{H}$ 
and T$_{max}^{L}$, corresponding to the heavy and light fermion species, respectively. 
T$_{max}^{L}$ and T$_{max}^{H}$ sets the scale for the regime of species dependent pseudogap 
behavior. 
We show these thermal scales in the $\eta$-T phase diagram in Fig.9.
There are two pseudogap regions as PG-I and PG-II. While within the PG-I regime 
both the fermion species are pseudogapped, it is only the light species which is pseudogapped in the 
PG-II regime. As $\eta \rightarrow$ 1 both the scales T$_{max}^{H}$ and T$_{max}^{L}$ collapses into
a single one, as expected from a mass balanced system. 

In the LO phase, mass imbalance gives rise to intriguing features both in the thermal and quasiparticle behavior.
 For a particular choice of population imbalance the mass 
imbalance ratio $\eta$ dictates the pairing momenta ${\bf Q}$. The signature of the same can be 
observed both in the pairing field structure factor (S(${\bf q}$)) as well as in pair correlation 
$\Gamma$(${\bf q}$), both of which shows peak at ${\bf q} \neq 0$.
We show the thermal evolution of the same in Fig.10 at $\eta = 0.6$ 
and a representative population imbalance of 
h=1.0t$_{L}$, corresponding to the LO phase.
At this choice of parameters the underlying LO phase is uniaxially modulated 
as can be seen from the two-fold symmetry of S(${\bf q}$) and 
$\Gamma$(${\bf q}$) at the lowest temperature. With T$_{c} \sim 0.01t_{L}$
we find that the state undergoes thermal disordering and acquires a four-fold
symmetry T$\approx$T$_{c}$. Short range LO pair correlations however continues 
to survive upto still higher temperatures and vanishes only at T$>$2T$_{c}$.
 
We next show how the finite momentum pairing in the LO regime modifies the
quasiparticle behavior. As mentioned above the DOS at the shifted Fermi level 
is pseudogapped even at the ground state and contains additional van Hove 
singularities. Thermal disordering smears out these singularities.
The exact number and location (energy)
of the van Hove singularities are altered by the choice of $\eta$. We demonstrate this behavior in 
Fig.11 where we show the thermal evolution of the species resolved DOS at two different $\eta$'s 
mentioned in the figure caption. At $\eta$=0.4 and 0.6, the finite momentum pairing takes place 
at ${\bf Q}$ = (0, $\pi$) and (0, $\pi$/2), respectively. At T=0, the DOS is computed using the 
variational scheme on a system size of 60 $\times$ 60. We have further compared our ground 
state results with the one obtained by Green's function formalism (shown by dotted curve) and
have observed qualitative agreement between the two.

%%%%%%%%%%%%%%%%%%%%%%%%%%%%%%%%%%%%%%%%%%%%%%%%%%%%%%%%%%%%%%%%%%%%%%%%%%%%
\begin{figure}
\begin{center}
\includegraphics[height=6.5cm,width=7.5cm,angle=0]{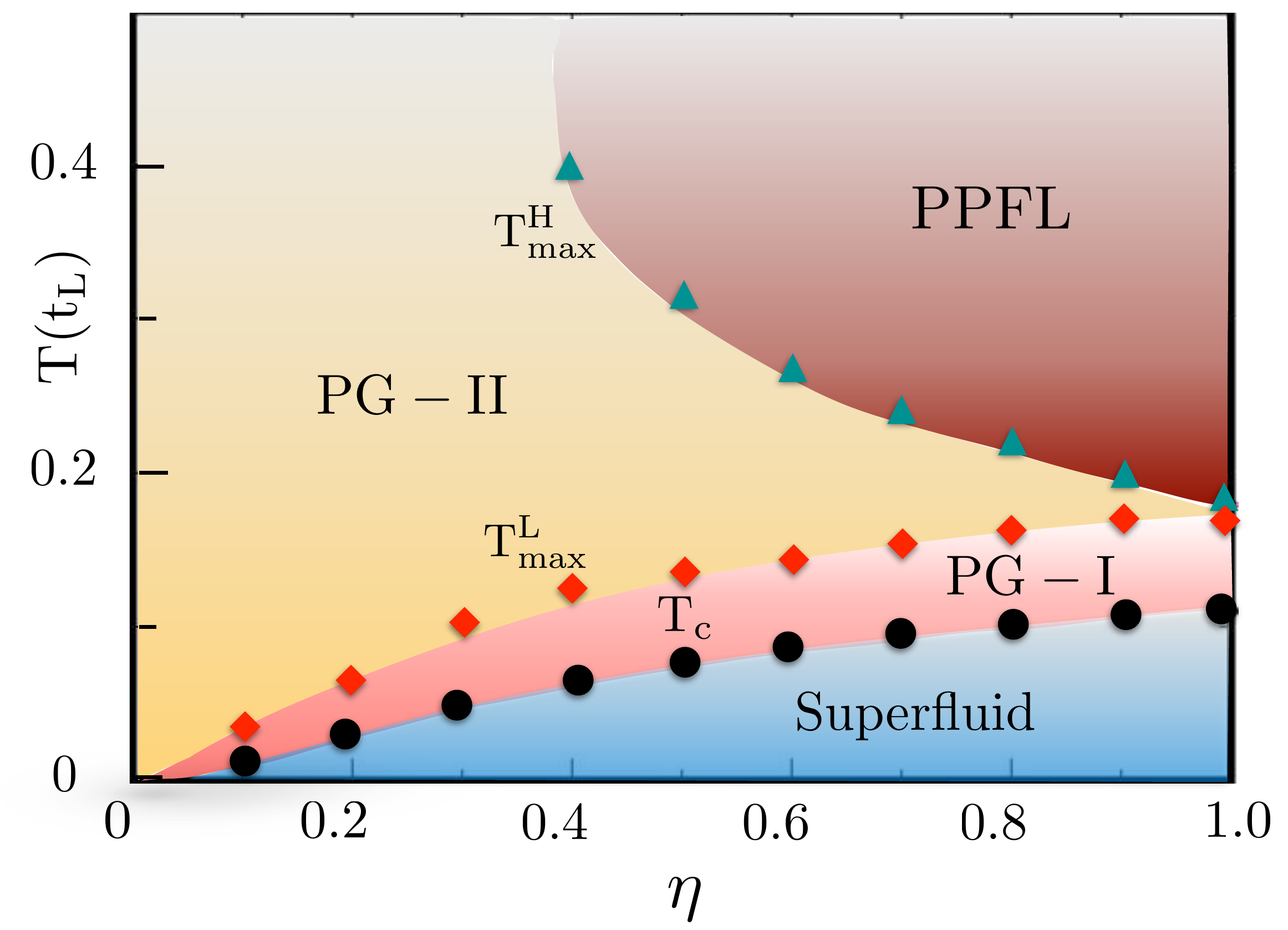}
\end{center}
\caption{Color online: Mass imbalance-temperature ($\eta$-T) phase diagram
at fixed population imbalance of $h=0.6t_{L}$. Along with the superfluid 
regime the figure shows the pseudogap regimes based on the species 
resolved DOS. T$_{max}^{L}$ and T$_{max}^{H}$ correspond to the scales 
beyond which the pseudogap behavior becomes non monotonic (see text).
PG-I corresponds to the regime where both the light and heavy species 
are pseudogapped, while in the PG-II regime only the light species 
is pseudogapped.}
\end{figure}  
%%%%%%%%%%%%%%%%%%%%%%%%%%%%%%%%%%%%%%%%%%%%%%%%%%%%%%%%%%%%%%%%%%%%%%%%%%%%

Thermal evolution of the species resolved spectral function A$_{\sigma}$(${\bf k}, \omega$) along the $\{0,0\}$ to $\{\pi,\pi\}$ scan across the Brillouin zone 
for two different 
choices of $\eta$ are shown next, in Fig.12. The multiband nature of the dispersion spectra 
is evident at low temperatures. In agreement with the DOS there is no hard 
gap at the Fermi level, while soft gaps or depletion of spectral weights 
are observed at the shifted Fermi levels.
As in case of the BP phase the thermal disordering scales continues 
to be species dependent. 
%%%%%%%%%%%%%%%%%%%%%%%%%%%%%%%%%%%%%%%%%%%%%%%%%%%%%%%%%%%%%%%%%%%%%%%%%%%
\begin{figure*}
\begin{center}
\includegraphics[height=6.5cm,width=15.5cm,angle=0]{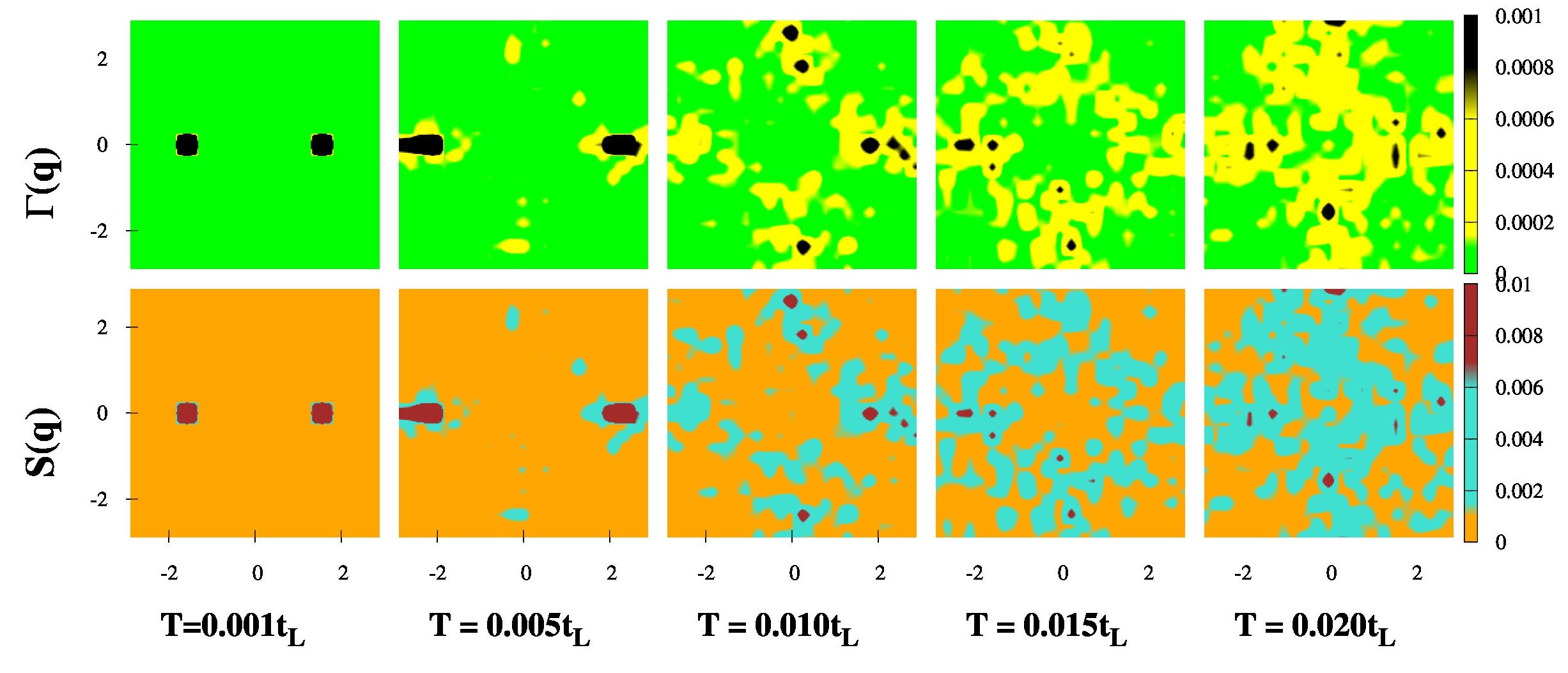}
\end{center}
\caption{Color online: Thermal evolution of pairing field structure 
factor (S(${\bf q}$)) and pair correlation ($\Gamma({\bf q})$) at a representative point 
(h=1.0t$_{L}$) in the LO phase 
for a mass imbalance ratio of $\eta$=0.6 and interaction U=4t$_{L}$. 
The underlying uniaxially modulated LO state is observed through the finite
${\bf Q}$ peaks in S(${\bf q}$) and $\Gamma$(${\bf q}$), at low temperatures.
Fluctuation progressively disorders the system and restore the four-fold 
symmetry. However, signatures of short range correlations survive upto 
T$>$2T$_{c}$ (where, T$_{c} \sim 0.01$t$_{L}$).}
\end{figure*}
%%%%%%%%%%%%%%%%%%%%%%%%%%%%%%%%%%%%%%%%%%%%%%%%%%%%%%%%%%%%%%%%%%%%%%%%%%%

Before closing this section we discuss about two quantities which crucially 
depends on the imbalance of population and mass in the system. 
The first quantity is the momentum resolved occupation of the fermion species 
n$_{\sigma}({\bf k})$, which maps out the Fermi surface architecture. In presence 
of an underlying inhomogeneous pairing field such as the LO superfluid we expect 
a non trivial Fermi surface and show the same for two different choices of mass 
imbalance ratio $\eta$=0.4 and 0.6, in Fig.13. As in case of DOS the T=0 
calculations are carried out on a larger system size of 60$\times$60 using 
the variational technique.  
Apart from the mismatch 
in size the Fermi surfaces now show two-fold symmetry consequent to the uniaxial modulation 
of the pairing field at this particular parameter point. Thermal evolution progressively smears 
out the directional asymmetry in the Fermi surface, and at T $>$ 2T$_{c}$ the expected 
four-fold symmetry is restored.     
Owing to the asymmetry of the Fermi surface, pairing now essentially takes
place only at selected ${\bf Q}$ values \cite{mpk2016}.  

The second quantity of interest is the low energy spectral weight distribution 
A(${\bf k}$, 0) which gives information about the nature of the superfluid 
gap. In Fig.14 we plot A(${\bf k}$, 0) at $\eta$=0.4 and 0.6 as it 
evolves in temperature. A very interesting behavior emerges from this figure, 
wherein in spite of an isotropic s-wave symmetry of the pairing field, the 
superfluid gap is now ``nodal'', arising purely out of the finite momentum scattering
that takes place in the LO phase. The gap isotropy is restored at T$>$2T$_{c}$.
Momentum resolved photoemission spectroscopy is one such experimental tool which 
can probe the angular dependence of the gap. While the presence of an 
underlying LO phase guarantees a nodal gap structure, the exact symmetry of 
the gap (as well as the Fermi surface) is dictated by the pairing momentum 
${\bf Q}$ and thus by the mass imbalance ratio $\eta$. We believe that such nontrivial 
behavior of the gap would have intriguing signatures in species resolved
transport measurements. We however do not touch upon those 
issues in the present paper.
%%%%%%%%%%%%%%%%%%%%%%%%%%%%%%%%%%%%%%%%%%%%%%%%%%%%%%%%%%%%%%%%%%%%%%%%%%%%%%%
\begin{figure}
\begin{center}
\includegraphics[height=7cm,width=8cm,angle=0]{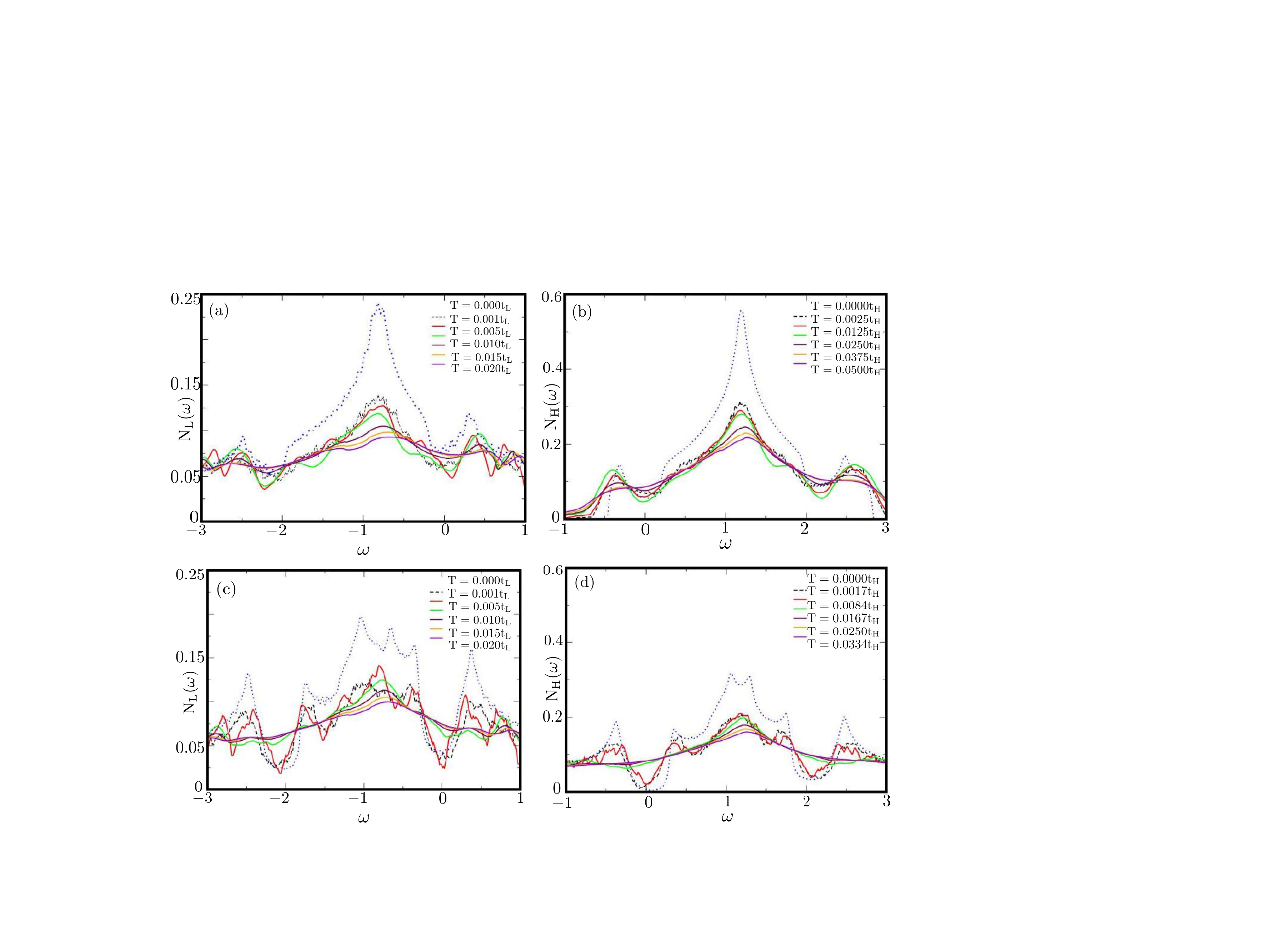}
\end{center}
\caption{Color online: Thermal evolution of species resolved DOS at a
population imbalance of 
h=1.0t$_{L}$ corresponding to the LO regime, for selected mass imbalance 
ratio of $\eta$=0.4 (panels (a)-(b)) and $\eta=0.6$ (panels (c)-(d)).
The state is pseudogapped even at the lowest temperature owing to the 
band structure effects. Note the additional van Hove singularities 
that arises due to the underlying modulated state with ${\bf Q} = \{0, \pi\}$
at $\eta$ = 0.4 and ${\bf Q} = \{0, \pi/2\}$ at $\eta$=0.6. The T=0 
DOS is determined from the variational calculation on a lattice size of 
60$\times$60, while for T $\neq$ 0 we use Monte carlo simulations on 
a lattice of size 24$\times$24. The dotted blue curve in each panel 
correspond to the results obtained by the Green's function formalism, 
at the ground state. Access to large system sizes by this technique 
enables us to demonstrate the van Hove singularities prominently.
The results obtained by Greens function formalism agrees reasonably 
with the one obtained through the variational calculations.}
\end{figure}
%%%%%%%%%%%%%%%%%%%%%%%%%%%%%%%%%%%%%%%%%%%%%%%%%%%%%%%%%%%%%%%%%%%%%%%%%%%%%%%
%%%%%%%%%%%%%%%%%%%%%%%%%%%%%%%%%%%%%%%%%%%%%%%%%%%%%%%%%%%%%%%%%%%%%%%%%%%%%                                      
\begin{figure*}
\begin{center}
\centerline{
\includegraphics[height=7.5cm,width=9.5cm,angle=0]{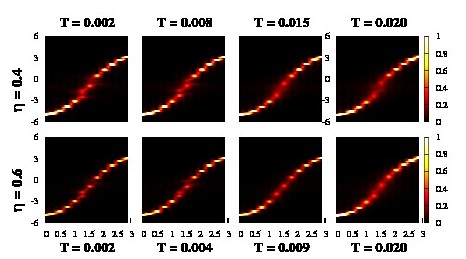}
\includegraphics[height=7.5cm,width=9.5cm,angle=0]{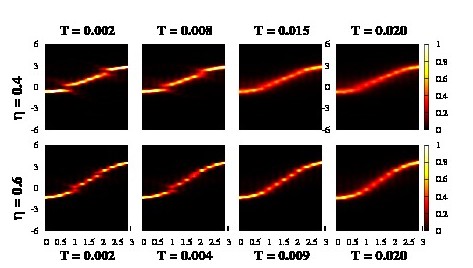}
}
\end{center}
\caption{Color online: Thermal evolution of light (left) and heavy (right) species
LO spectral function A(${\bf k}, \omega$) across the Brillouin zone 
for ${\bf k} = \{0, 0\}$ to $\{\pi, \pi\}$ at h=1.0t$_{L}$, U=4t$_{L}$ and mass 
imbalance ratio $\eta$ = 0.4 and 0.6. Note the multi-branch nature 
of the dispersion spectra arising due to LO modulations. Note that there is no hard gap at the 
Fermi level, rather there is depletion of spectral weight at the shifted Fermi level 
($\omega = \pm h$).}
\end{figure*}
%%%%%%%%%%%%%%%%%%%%%%%%%%%%%%%%%%%%%%%%%%%%%%%%%%%%%%%%%%%%%%%%%%%%%%%%%%%%%                                      
\subsection{Effect of interaction}

In the last few sections we have discussed how the interplay of mass and population imbalances
bring about several intriguing features in a Fermi-Fermi mixture, at a particular interaction 
strength. One of the principal advantages of the cold atomic gas quantum emulator 
is the ability to control the interaction strength and thus it is of significant interest 
to understand how the interplay between the population and mass imbalances alter the 
well known picture of BCS-BEC crossover in balanced Fermi gas. 

In this section we briefly discuss the interplay and present our observations in 
terms of the m-T phase diagram at $\eta$ = 0.6 for different choices of interaction 
strength in Fig.15. The phase diagram remains qualitatively the same at other mass 
imbalance ratio. 
Fig.15 can roughly be compared with Fig.8 and 9 of ref.\cite{stoof2010}. 
In the limit of small polarization the system is in the breached pair state 
comprising of uniform superfluidity with finite polarization. 
Note that we do not make a distinction between a BCS state with a gap at the Fermi level 
and a Sarma phase with gapless superconductivity, as has been discussed in ref.\cite{stoof2010}.
At the interaction regime we are in the BCS description of the state ceases to be valid.
At T$\neq$0 there is spontaneous emergence of islands with non-zero polarization, 
giving rise to coexisting superfluid and magnetic behavior in the BP phase \cite{mpk2016}.
At the tricritical point T$_{c1}$ the order of thermal transition changes from second 
to first {\it within the BP phase}. Akin to the continuum case \cite{stoof2010} the 
first order transition is marked by discontinuity is polarization as well as in density.
The resulting forbidden region in Fig.8 and 9 of ref.\cite{stoof2010} is the 
unstable region in Fig.15, of the present paper. At still larger polarization a 
first order transition takes place between 
the unstable BP phase and the LO phase at the Lifsitz point T$_{c2}$. However, it must be 
noted that in case of lattice fermions for weak and intermediate interactions  the T$_{c1}$
and T$_{c2}$ are distinct, unlike the continuum phase diagram where the Lifshitz point
coincides with the tricritical point. As shown in Fig.15, at strong interaction (U=6t$_{L}$)
there is indeed a single critical point where the BP 
phase undergoes a second order transition to an LO phase, without an intervening first order
unstable regime.   

%%%%%%%%%%%%%%%%%%%%%%%%%%%%%%%%%%%%%%%%%%%%%%%%%%%%%%%%%%%%%%%%%%%%%%%%%%%%%%%%%%               
\begin{figure*}
\begin{center}
\includegraphics[height=8.5cm,width=15.5cm,angle=0]{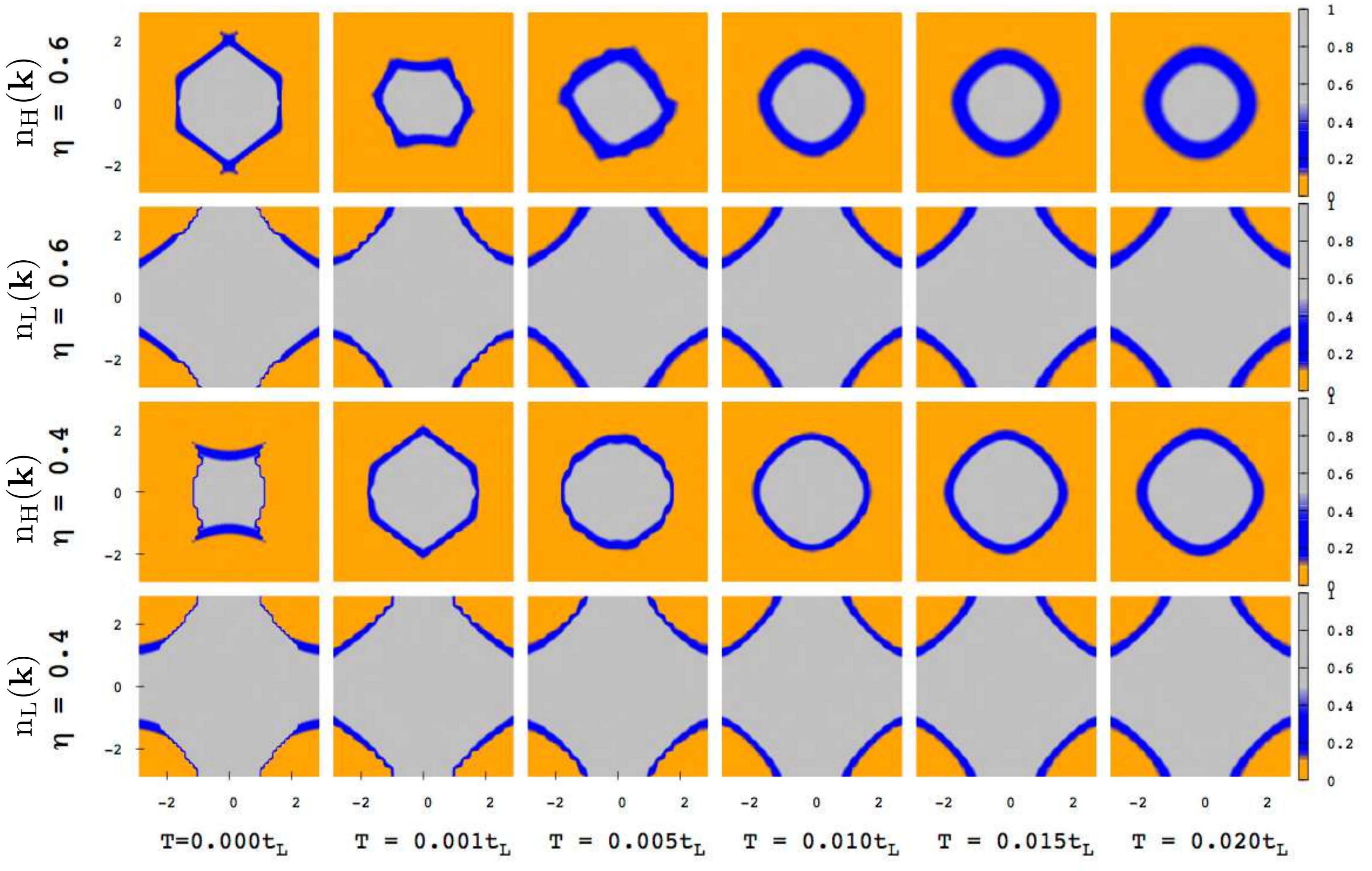}
\end{center}
\caption{Color online: Thermal evolution of single particle occupation for the different
species (n$_{\alpha}({\bf k})$), where $\alpha = L,H$; mapping out the
Fermi surface at $\eta$=0.4 and 0.6 for h=1.0t$_{L}$. Note the mismatch in size
of the Fermi surfaces owing to the imbalance. The T=0 results are once again obtained
using the variational calculation at large system size of L=60.
The anisotropy in the Fermi surface architecture arises due to the
modulated underlying state. The isotropy of the Fermi surface is regained
at high temperature.}
\end{figure*}
%%%%%%%%%%%%%%%%%%%%%%%%%%%%%%%%%%%%%%%%%%%%%%%%%%%%%%%%%%%%%%%%%%%%%%%%%%%%%%%%%%         
%%%%%%%%%%%%%%%%%%%%%%%%%%%%%%%%%%%%%%%%%%%%%%%%%%%%%%%%%%%%%%%%%%%%%%%%%%%%%%%%%%               
\begin{figure*}
\begin{center}
\includegraphics[height=6cm,width=16cm,angle=0]{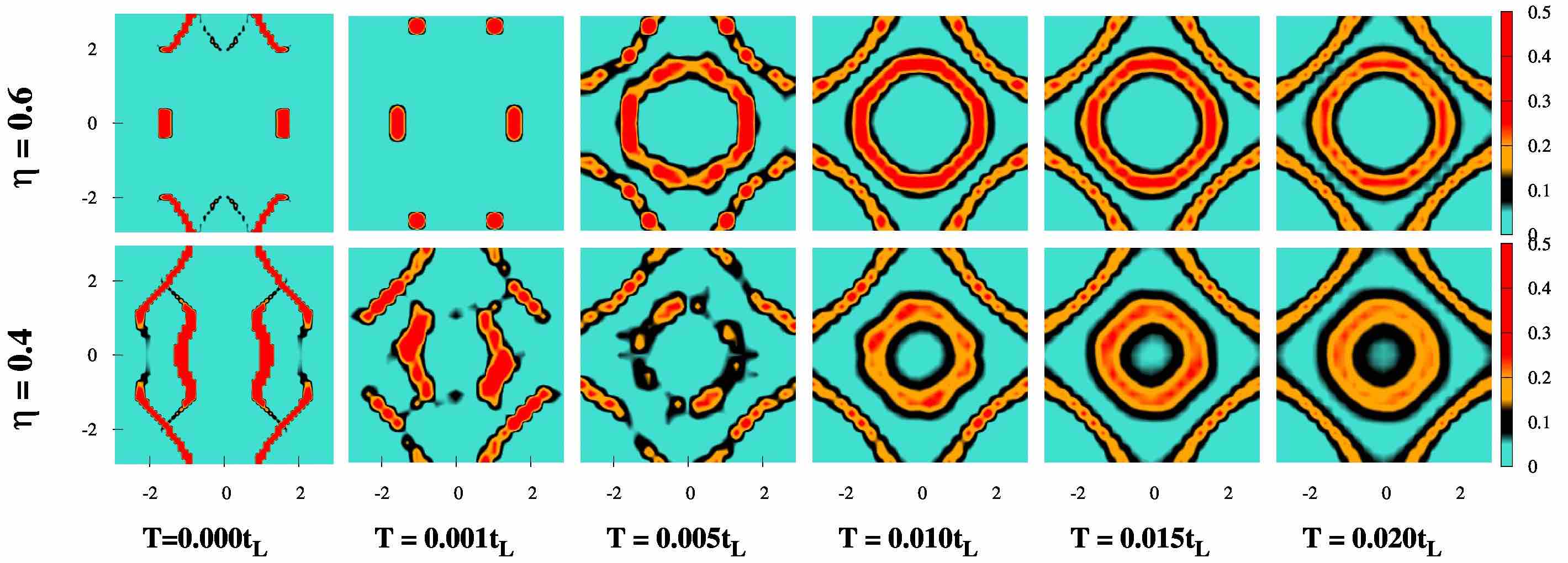}
\end{center}
\caption{Color online: Thermal evolution of low energy spectral
weight distribution A(${\bf k}, 0$) at the Fermi level $\eta$=0.4 and 0.6,
mapping  out the superconducting gap structure. Note that inspite 
of an isotropic s-wave pairing field a ``nodal'' gap structure is
realized at low temperature. At T$\sim$2T$_{c}$ the gap isotropy is
restored.}
\end{figure*}
%%%%%%%%%%%%%%%%%%%%%%%%%%%%%%%%%%%%%%%%%%%%%%%%%%%%%%%%%%%%%%%%%%%%%%%%%%%%%%%%%%    
%%%%%%%%%%%%%%%%%%%%%%%%%%%%%%%%%%%%%%%%%%%%%%%%%%%%%%%%%%%%%%%%%%%%%%%%%%%%%%%%%%%%%%%%
\begin{figure*}
\begin{center}
\includegraphics[height=6cm,width=15cm,angle=0]{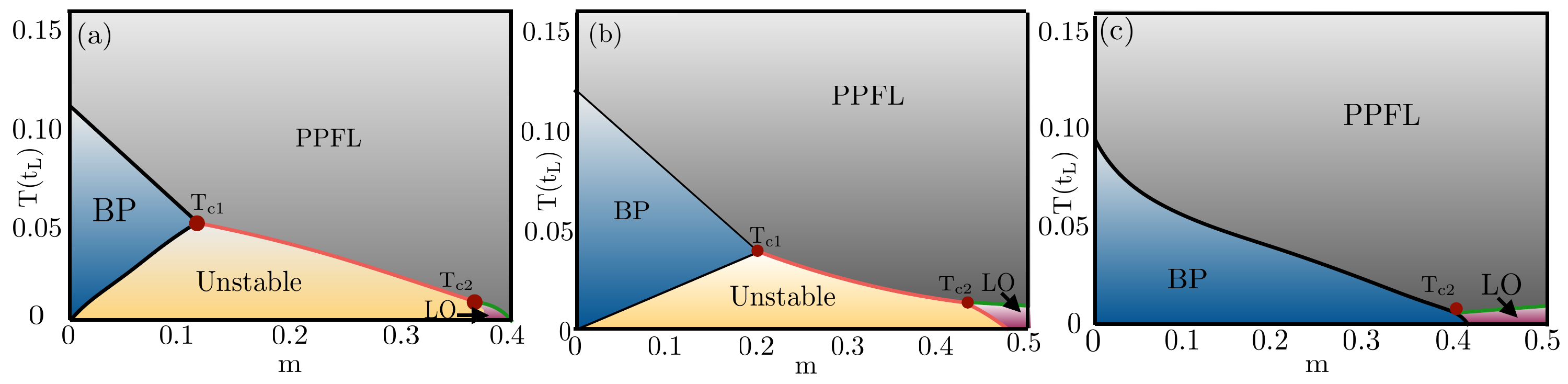}
\end{center}
\caption{Color online: Polarization-temperature (m-T) phase diagram 
for $\eta$=0.6 at different interactions (a) U=3t$_{L}$, (b) U=4t$_{L}$
and (c) U=6t$_{L}$. At strong interaction (U=6t$_{L}$) the system undergoes a second 
order transition from a BP to LO regime ``without'' an intervening 
unstable regime involving first order transition.}
\end{figure*}  
%%%%%%%%%%%%%%%%%%%%%%%%%%%%%%%%%%%%%%%%%%%%%%%%%%%%%%%%%%%%%%%%%%%%%%%%%%%%%%%%%%%%%%%%

\subsection{Comparison with experiments}

We now consider an experimentally realizable Fermi-Fermi mixture and attempt 
to make some quantitative predictions about it based on our discussions in the
previous sections. A suitable candidate for the same is 6$_{Li}$-40$_{K}$ 
mixture \cite{taglieber2008, naik2010, wille2008, costa2010, naik2010}.
Even though superfluidity
is yet to be achieved, Feshbach resonance as well as the formation of 
heteromolecules 6$_{Li}$-40$_{K}$ has already been attained for this mixture. 

In order to analyze the behavior of 6$_{Li}$-40$_{K}$ mixture
we choose the mass imbalance ratio to be $\eta$=0.15. For the population imbalance 
we select h=0.6t$_{L}$ corresponding to the system in BP regime, close to unitarity. 
We believe that as compared to the LO superfluid regime the BP phase is more readily 
accessible to the experimental probes, owing to it's higher thermal scales. 
This justifies our choice of h=0.6t$_{L}$.

We begin the discussion of our results by demonstrating the BCS-BEC crossover for 
the 6$_{Li}$-40$_{K}$ mixture.
Fig.16a shows the pairing field structure factor 
(S(0,0)) (at ${\bf q}$ = 0) across the BCS-BEC cross over. In the intermediate and
strong coupling regime (U $\ge$ 4t$_{L}$) the ground state of the system at this
population imbalance correspond to an unpolarized superfluid (USF). The finite
temperature counterpart of the same leads to the BP phase. In the weak coupling
regime (U$<$ 4t$_{L}$) the system is a partially polarized Fermi liquid (PPFL) in the ground state and does not show any long range order.
%%%%%%%%%%%%%%%%%%%%%%%%%%%%%%%%%%%%%%%%%%%%%%%%%%%%%%%%%%%%%%%%%%%%%%%
\begin{figure}
\begin{center}
\includegraphics[height=7.2cm,width=9cm,angle=0]{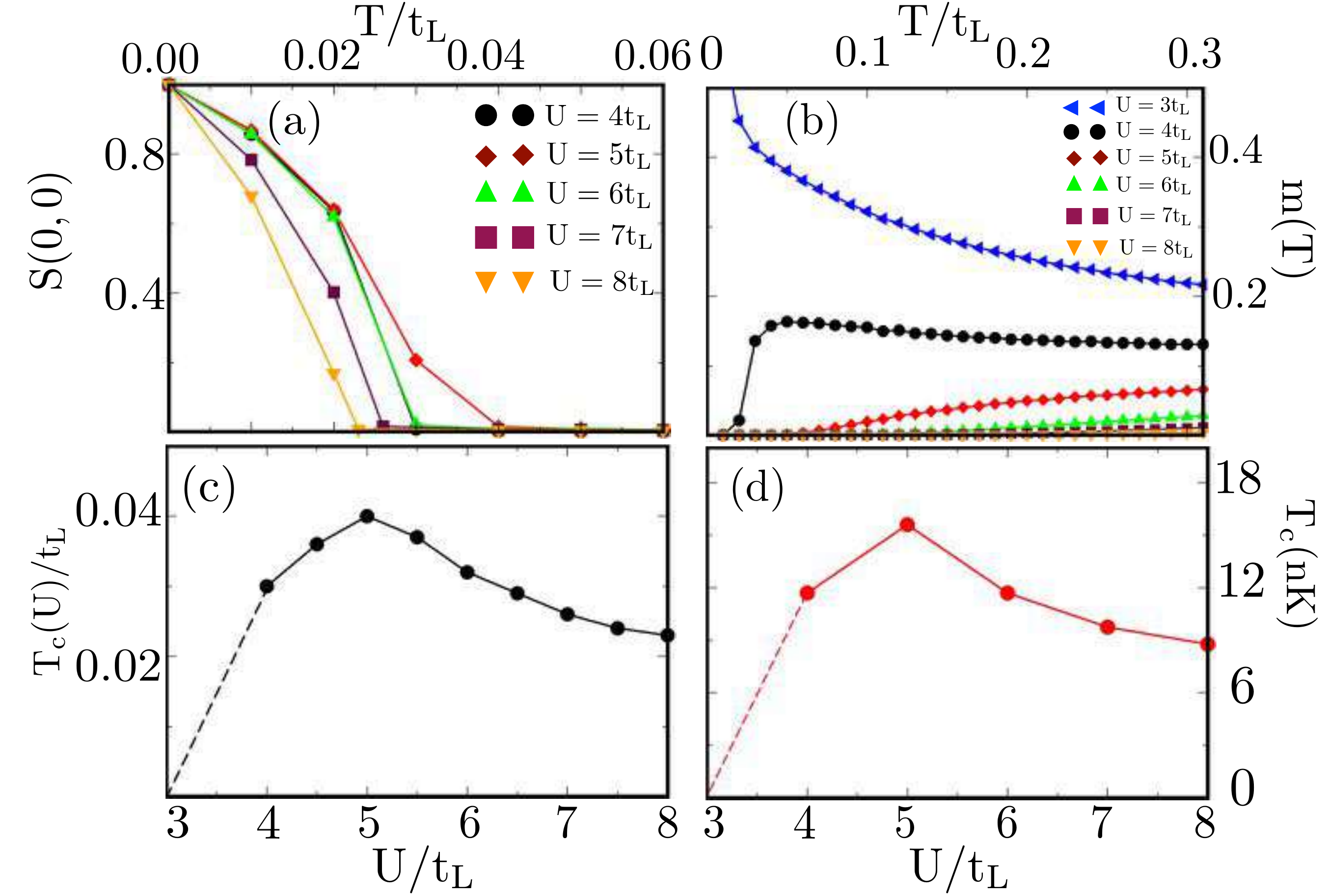}
\end{center}
\caption{Color online: (a) Pairing field structure factor (S(0, 0))
 at a population imbalance of h = 0.6t$_{L}$ and mass imbalance ratio 
 of $\eta$ = 0.15 (corresponding to the 6$_{Li}$-40$_{K}$ mixture) 
for different interactions. (b) Temperature 
 dependence of polarization (m(T)) at different interactions.
  A large interaction suppresses the polarization. 
 (c) BCS-BEC crossover at $\eta$ = 0.15.
 T$_{c}$(U) has its maxima (corresponding to unitarity) at U = 5t$_{L}$,
 (d) BCS-BEC crossover replotted in terms of the experimental scales 
appropriate for the 6$_{Li}$-40$_{K}$ mixture. Note that at a fixed
population imbalance the uniform superfluidity sets in beyond a
critical interaction (U $>$ U$_{c}$). The dashed line shows that for 
U $<$ U$_{c}$ the superfluid order collapses.}
\end{figure}
%%%%%%%%%%%%%%%%%%%%%%%%%%%%%%%%%%%%%%%%%%%%%%%%%%%%%%%%%%%%%%%%%%%%%%%

Fig.16b shows the temperature dependence of polarization across the BCS-BEC cross over. 
Increasing interaction suppresses the polarization and uniform superfluid state is
realized over a wider regime of temperature. At large interactions where the fermions
form tightly bound pairs, the required population imbalance to break the pair and create
finite polarization is large, leading to the suppression in the polarization. In striking
contrast is the weak (U $<$ 4t$_{L}$) interaction limit where even at the ground state
there is a large finite polarization, indicating a PPFL state.
  
We present the T$_{c}$ scale for the 6$_{Li}$-40$_{K}$ mixture as 
determined from S(0, 0), in Fig.16c. There are two key effects which
decides the behavior of the T$_{c}$ scale. We discuss them point-
wise. (i) The primary effect is the suppression of the T$_{c}$ by the
imbalance in mass. Close to unitarity, at U = 4t$_{L}$ the T$_{c}$ for a
mass balanced system is T$_{c}^{bal}$ $\sim$ 0.15t$_{L}$, in comparison to 
T$_{c}^{imb}$ $\sim$ 0.03t$_{L}$ $\sim$ 0.2T$_{c}^{bal}$ for 6$_{Li}$-40$_{K}$
mixture. The estimated T$_{c}$
of the 6$_{Li}$-40$_{K}$ mixture (at U=4t$_{L}$) amounts to T$_{c}^{imb}$ $\sim$ 11.7nK.
(ii) The second effect on T$_{c}$ arises out of the population imbalance.
An weaker interaction shrinks the regime of both the uniform and modulated
superfluid and rapidly gives way to a PPFL state. Thus, at a fixed population 
imbalance the system can be in a PPFL state at weak interactions, while a large
interaction would correspond to an uniform superfluid state at the same imbalance. 
Consequently, at a fixed population imbalance, on traversing through the BCS-BEC crossover an
uniform superfluid state would be realized only beyond a critical interaction (U$_{c}$).
The behavior is significant and is in contrast to the balanced system in which any 
arbitrarily small attractive interaction gives rise to an uniform superfluid state.

Fig.16c further shows that T$_{c}$(U) has a peak at U = 5t$_{L}$, corresponding to the
unitarity in the context of lattice fermion model (see discussion section).
We estimate the T$_{c}$(U) scale in
experimental units and map out the expected BCS-BEC crossover for the 6$_{Li}$-40$_{K}$ 
mixture in Fig.16d. At unitarity the mixture has a T$_{c}$ $\sim$ 15nK.

In Fig.17 we show the thermal evolution of DOS for the two species at different interactions.
At U = 3t$_{L}$ there is no long range ordered ground state of the system, consequently
there is no hard gap at the Fermi level even at the lowest temperature. There is a depletion 
in spectral weight at the Fermi level, giving rise to a “pseudogap” phase. Thermal evolution
leads to further depletion of the spectral weight at the Fermi level, in agreement with m(T) 
(Fig.16b) which shows a reduction in polarization at high temperatures, indicating emergence
of short range correlations. A larger interaction (U = 4t$_{L}$) gives rise to a hard gap at
the Fermi level, which fills up monotonically with temperature before undergoing a non
monotonic thermal evolution at T$^{L}_{max}$ (T$^{H}_{max}$) corresponding to the light
(heavy) species. At U=5t$_{L}$ the T$^{L}_{max}$ and T$^{H}_{max}$ scales are
significantly high and are not shown in Fig.17.
%%%%%%%%%%%%%%%%%%%%%%%%%%%%%%%%%%%%%%%%%%%%%%%%%%%%%%%%%%%
\begin{figure*}
\begin{center}
\includegraphics[height=8cm,width=14cm,angle=0]{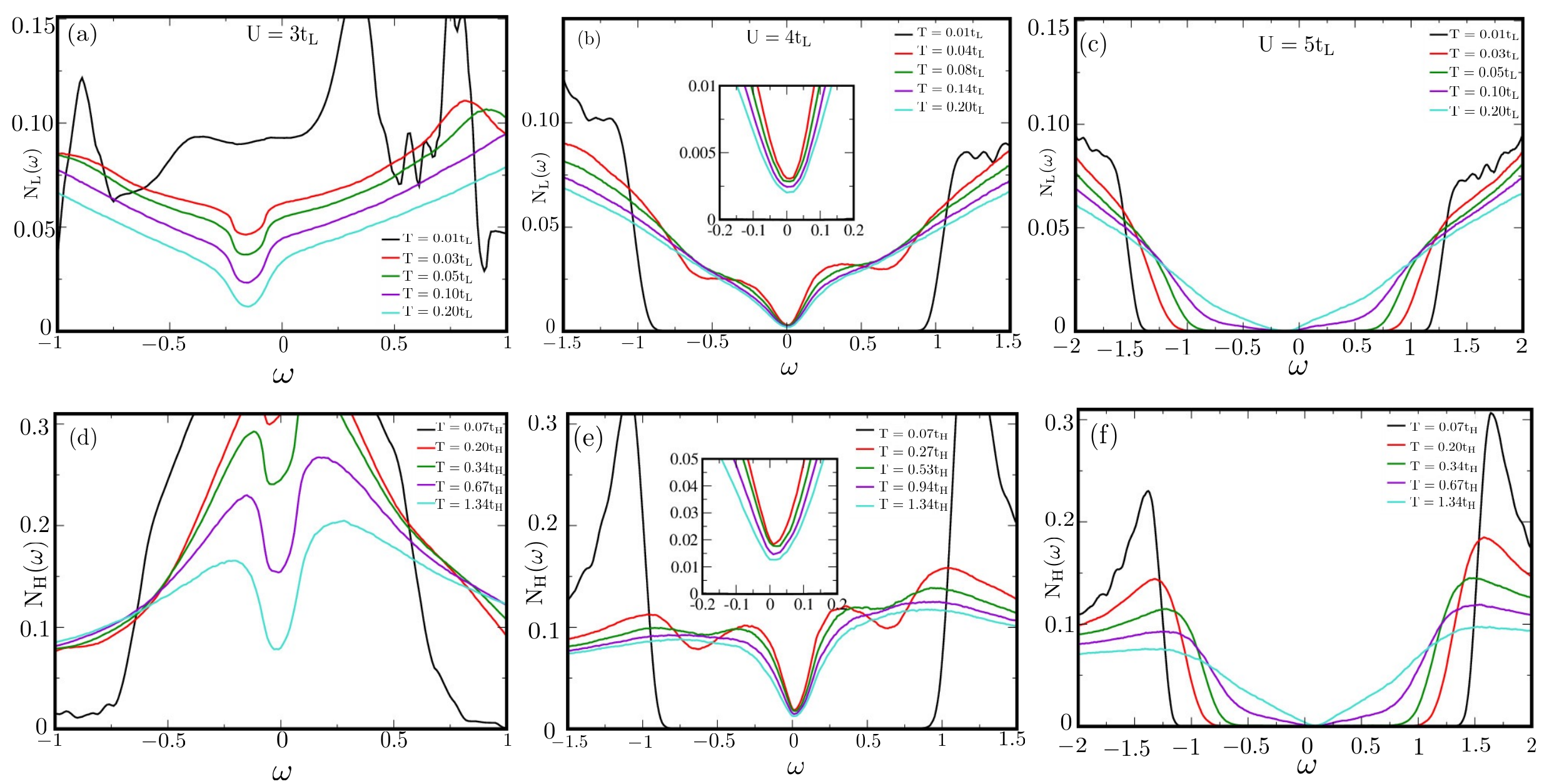}
\end{center}
\caption{Color online: Density of states (DOS) corresponding to the
light (a-c) and heavy (d-f) fermion species for a mass imbalance 
ratio of $\eta$ = 0.15 for selected U-T cross sections. The 
temperatures corresponding to the two species are normalized by 
their respective kinetic energy scales. At U = 3$t_{L}$, in the
temperature regime under consideration the DOS pertaining to both
species shows nonmonotonic thermal evolution. There is no hard 
gap at the Fermi level at this interaction since the 
lowest temperature state correspond to a partially polarized Fermi liquid 
(PPFL). At U = 4$t_{L}$ the DOS exhibits non monotonic behavior with 
increasing temperature and leads to depletion of the spectral weight at the 
 Fermi level. The onset of non monotonicity marks the temperature scales 
 T$_{max}^{H}$ and T$_{max}^{L}$ (see text) corresponding to the heavy and 
 light species respectively. The inset highlights the nonomonotonic 
 behavior close to the Fermi level. At U = 5t$_{L}$ the thermal 
 evolution of the 
 DOS is monotonic upto a very high temperature, beyond which the scales 
 T$_{max}^{L}$ and T$_{max}^{H}$ (not shown in the figure) sets in}
\end{figure*}
%%%%%%%%%%%%%%%%%%%%%%%%%%%%%%%%%%%%%%%%%%%%%%%%%%%%%%%%%%%

We next analyze the momentum resolved spectral function A$_{\sigma}$ (${\bf k}, \omega$).
In Fig.18 we show the thermal
evolution of the species resolved spectral function at three different interactions,
along the $\{0, 0\}$ to $\{\pi, \pi\}$ scan across the Brilllouin zone. The figure shows
species dependent thermal scales in the problem.

The spectral function for U=3t$_{L}$ reveals that at the lowest temperature there is
finite weight at the Fermi level and consequently the dispersion spectra is gapless 
for both the species. Increase in temperature leads to depletion of weight, giving
rise to a small but finite gap at the Fermi level. This is a special case of 
temperature driven gapless to gap transition, arising out of short range correlations.
%%%%%%%%%%%%%%%%%%%%%%%%%%%%%%%%%%%%%%%%%%%%%%%%%%%%%%%%%%%%%%%
\begin{figure*}
\centerline{
\includegraphics[height=7.5cm,width=7.5cm,angle=0]{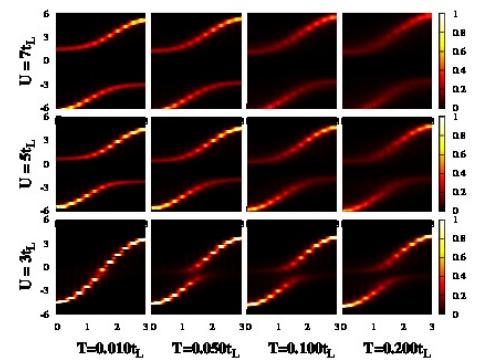}
\includegraphics[height=7.5cm,width=7.5cm,angle=0]{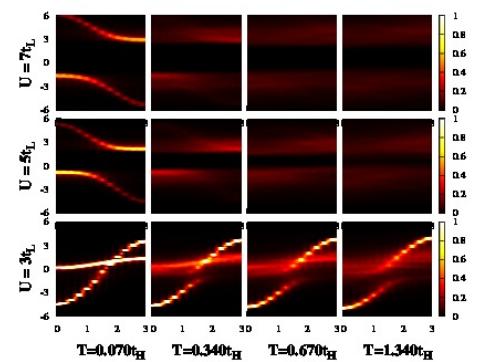}
}
\caption{Color online: Spectral function maps along the $\{0, 0\}$ to 
 $\{\pi, \pi\}$ scan across the Brilllouin zone for the (left) light and
 (heavy) heavy fermion species at $\eta$ = 0.15 and population imbalance
  of h=0.6t$_{L}$, 
 for different U - T cross sections. Notice that at U = 3$t_{L}$ increase 
 in temperature opens up a gap at the shifted Fermi level in the dispersion 
 spectra corresponding to both species. The heavy species undergoes faster 
thermal disordering since it experiences higher scaled temperature.}
\end{figure*}
%%%%%%%%%%%%%%%%%%%%%%%%%%%%%%%%%%%%%%%%%%%%%%%%%%%%%%%%%%%%%%%

We now present the thermal phase diagram
across the BCS-BEC crossover for the 6$_{Li}$-40$_{K}$ mixture, in Fig.19a. 
While in the PG-I regime both 6$_{Li}$ and 40$_{K}$ would show pseudogap 
behavior, in the PG-II regime it is only the 6$_{Li}$ species which is
in the pseudogap phase. 40$_{K}$ in the PG-II regime is a partially 
polarized Fermi liquid.
 Note that a similar observation
has also been made in the context of continuum model \cite{ohashi2013, ohashi2014}. 
Both T$_{max}^{L}$ and 
T$_{max}^{H}$ are significantly higher than T$_{c}$ and consequently are better 
accessible to experimental probes, such as rf spectroscopy. 
%%%%%%%%%%%%%%%%%%%%%%%%%%%%%%%%%%%%%%%%%%%%%%%%%%%%%%%%%%%%%%%%%%%%%%%%%%
\begin{figure}
\begin{center}
\includegraphics[height=6cm,width=9cm,angle=0]{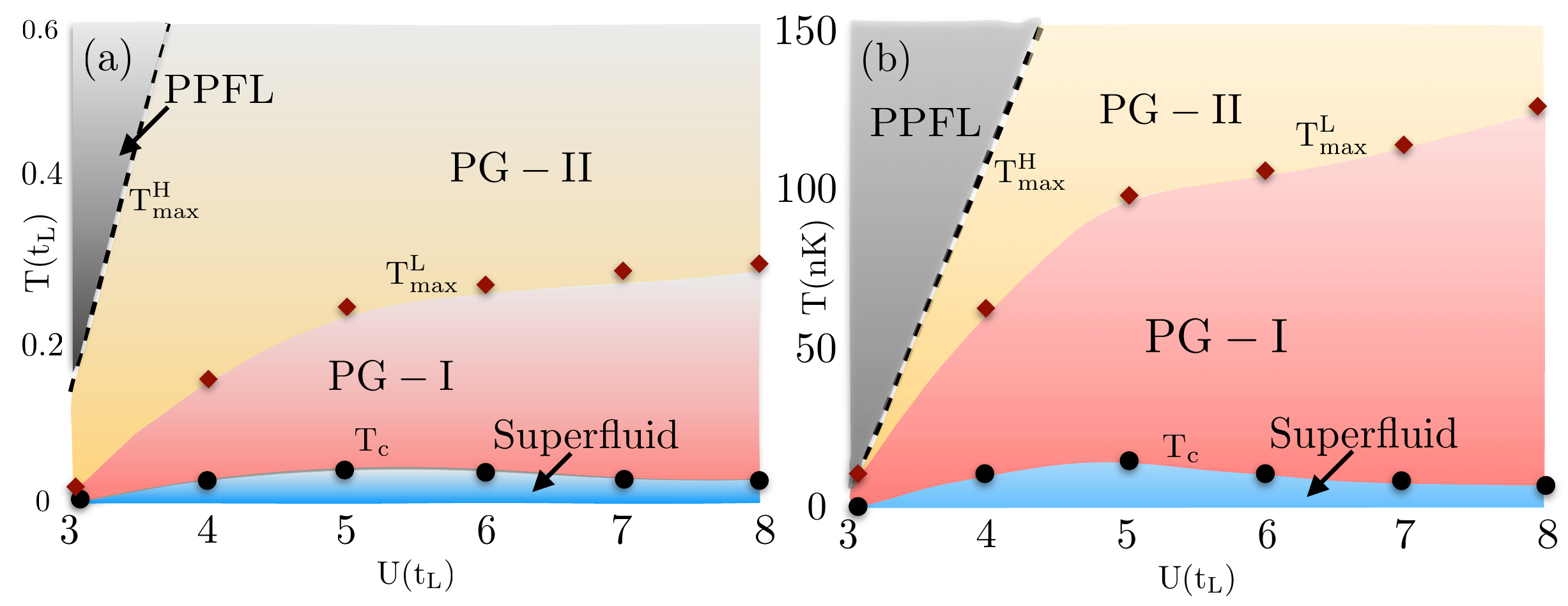}
\end{center}
\caption{Color online: (a) Interaction-temperature (U-T) phase diagram at 
 $\eta$ = 0.15 corresponding to the 6$_{Li}$-40$_{K}$ mixture.
 The figure shows the T$_{c}$ scale along with the 
 pseudogap scales for this mixture at a population imbalance of h = 0.6$t_{L}$.
 Both the species exhibit pseudogap behavior in the PG-I 
regime, while in the PG-II regime only the light
species is pseudogapped. The thermal scales of the individual 
species are defined w. r. t. their corresponding kinetic energy scales.
(b) Thermal scales in terms of the experimental units as would be observable
in 6$_{Li}$-40$_{K}$ mixture, in species resolved rf spectroscopy.
Close to unitarity (U = 4$t_{L}$),  the pseudogap phase is
expected to be observable in both the species upto T $\sim$ 58nK, 
beyond which the pseudogap feature survives only in the light species upto 
T $\sim$ 108nK. Thus, even though the T$_{c} \sim$15nk is strongly 
suppressed in this mixture, short range pair correlations should be 
observable upto significantly higher temperatures, in experiments.}
\end{figure}
%%%%%%%%%%%%%%%%%%%%%%%%%%%%%%%%%%%%%%%%%%%%%%%%%%%%%%%%%%%%%%%%%%%%%%%%%

For 6$_{Li}$-40$_{K}$
mixture close to unitarity we expect the PG-I regime to survive upto T$_{max}^{L}$ 
$\sim$ 58.5nK while the PG-II regime should be observable even at T$_{max}^{H}$ $\sim$ 
108nK, as shown in Fig.19b. Another suitable probe is the momentum resolved spectroscopy
which can provide evidence of unequal thermal disordering temperatures corresponding to 
the two fermion species. A behavior that would be qualitatively similar to the one shown
in Fig.18.

\section{Discussion}

The preceding sections comprise of the main results of this work. 
We now touch upon in brief certain aspects on the model under consideration 
and it's connection to cold atomic experiments in continuum; we also discuss about the 
Green's function formalism that has been used in this work to compute the 
ground state quasiparticle properties and benchmark the results obtained with those 
from Monte carlo simulations. 

\subsection{Connection to continuum unitary gas}

The results presented in this paper are based on a lattice fermion model while 
at the same time they are motivated by experiments on unitary Fermi gas. In this
regard there are few issues that needs highlighting. We discuss them below. 

\subsubsection{Concept of unitarity}

For the cold atomic gases the interaction strength is quantified in terms 
of the s-wave scattering length $a_{D}$, with $D$ being the spatial dimensionality.
The corresponding coupling constant is defined as $k_{F}a_{D}$, where $k_{F}$
is the Fermi wave vector. For a 3D gas, the limit of unitarity can be
defined as the 
coupling strength at which the first two-body bound state is formed. With $a_{3D} \rightarrow \infty$
as $g \rightarrow g_{c}$, where $g$ is the interaction strength, it can be easily seen that 
$1/k_{F}a_{3D} = 0$ at $g_{c}$, corresponding to unitarity. At the same time $g_{c}$
corresponds to the point across the BCS-BEC crossover where the transition temperature 
is maximum, with $T_{c}^{max}/E_{F} \sim 0.15$ for 3D Fermi gas. Within the framework of 
a lattice fermion model (3D Hubbard model), it has been found that the first two-body 
bound state is formed at a critical interaction strength of $U_{c}/t \sim 7.9$. 
Interestingly, quantum Monte carlo (QMC) studies have found that for a 3D system 
the maximum T$_{c}$ is at $U/t \sim 8$ \cite{beck}. This brings out the concept that the 
critical interaction for the formation of two body bound state in a lattice
fermion model coincides with the interaction at maximum $T_{c}$. 

In case of 2D gas at continuum $a_{2D} \rightarrow \infty$ as $g \rightarrow 0$, since 
the two-body bound state is formed at any arbitrary interaction. This definition however 
corresponds to deep inside the BCS regime where a weak coupling description is valid. 
With increasing interaction the system crosses over to Bose limit as $a_{2D} \rightarrow 0$.
The coupling at crossover is defined via $ln(k_{F}a_{2D}) \rightarrow 0$. 
Interpolation between the BCS and BEC limits showed that maximum T$_{c}$ occurs 
at $ln(k_{F}a_{2D}) \rightarrow 0$, with $T_{c}^{max}/E_{F} \sim 0.1$ \cite{jochim}.
While a 3D like definition ($a_{2D} \rightarrow \infty$ as $g \rightarrow 0$) puts the 
unitarity limit in 2D deep inside the BCS regime, an alternate definition as 
$[ln(k_{F}a_{2D})]^{-1} \rightarrow \infty$ is a better choice, firstly because 
it captures the crossover between the BCS and BEC regimes correctly and also because it 
corresponds to the maximum T$_{c}$. In other words, the definition $[ln(k_{F}a_{2D})]^{-1} 
\rightarrow \infty$ is adequate to capture the two important features of unitarity in 2D, viz.
(i) at unitarity neither a pure bosonic nor a fermionic description is sufficient and 
(ii) the T$_{c}$ is maximum.

QMC calculation on 2D Hubbard model have shown that the maximum T$_{c}$ is obtained at 
U/t $\sim 5$. In our numerical simulations we call U/t $\sim 4$ being close to unitarity
where T$_{c} \sim$ 0.9T$_{c}^{max}$ \cite{paiva2d}.

\subsubsection{Continuum limit out of lattice model}

The present calculations is carried out at a fixed total chemical potential of 
$\mu = -0.2t_{L}$, which is close to the half filling and the Fermi surface is distinctly non circular and the 
lattice effects are dominant. We observe that even in the limit of low density 
where the Fermi surface is circular and $\epsilon_{\bf k} \sim k^{2}$ is a 
reasonable approximation, it is 
difficult to capture the continuum effects through the lattice model. At
interactions close to unitarity (U $\sim$ 5t$_{L}$) even if the Fermi levels 
occupy the lower edge of the band, scattering effects couple the states at 
upper edge. With these high energy states being lattice specific, even at 
low enough densities the results obtained do not match with that of continuum 
\cite{castelleni, capone}. In order to obtain continuum ``universal''
 physics out of 
lattice simulations one needs to go to extremely low densities $\sim$ 0.001,
corresponding to lattice size of $\sim 10^{4}$. This is however outside the range 
of what can be attained in the present day.       

\subsection{Single channel decomposition}

In the present work a single-field decomposition in the
pairing channel has been used. In general such a decomposition 
can not capture the instabilities in all the channels and one 
needs to take into account the decomposition in the pairing, 
density and spin channels,  particularly in the FFLO regime. 
However, in one of our recent works \cite{mpk2016} we have 
shown that even in the FFLO phase the density channel modulations 
are very weak. Moreover, being away from half filling the 
density modulations are not as important in the present study 
as it would have been at n=1. Decomposition in the additional 
magnetic channel might lead to quantitative difference in our 
results. However, while such multi-channel decomposition can be 
readily incorporated within a mean field formalism, a non 
gaussian fluctuation theory like the one presented in this work would 
be a difficult feat to achieve with a multi-channel decomposition. 
In order to keep the problem numerically tractable we have chosen 
for a single channel decomposition. We believe that inclusion 
of other channels would not lead to qualitative changes in our results. 

\subsection{Effect of quantum fluctuations}

One of the principal approximations that has been used in this 
work is the neglect of the quantum fluctuations. As discussed 
earlier we treat the pairing field as classical and retain the 
spatial fluctuations while neglecting the temporal fluctuations.
Within the framework of continuum model, this could be a poor 
approximation however in case of a lattice model it is reasonable.
In the continuum FFLO state the low energy fluctuation arises 
from, (i) the phase symmetry of the U(1) order parameter, 
(ii) the translational and (iii) the rotational symmetry 
breaking \cite{radz}. Consequently, in two-dimensional system, 
long range order can not be sustained even at T=0, rendering 
the corresponding mean field theory invalid. 
In a lattice model, while the phase field has ``XY'' type 
low energy excitations, the translational and rotational modes
are already gapped out since the spatial symmetry is 
already broken by  the underlying lattice \cite{loh2010}.
For example, it is well known that models with XY symmetry shows long range
order in 2D and undergoes BKT transition at finite temperature.
The issue of fluctuation thus reduces to verifying how well the U(1)
symmetry T$_{c}$ is captured by our model in comparison to a 
full QMC study. A population imbalanced system is difficult 
to be studied within a QMC approach owing to the fermionic sign 
problem. However, benchmarking the results for a balanced system 
as obtained by our technique with those obtained using QMC shows
fairly good agreement \cite{tarat_epl}. The comparison along with 
the arguments presented above suggests that our technique is 
suitable to capture the relevant fluctuations and the 
corresponding finite temperature behavior.   

\subsection{Finite size effect}

We have shown that unlike the balanced system, in presence of population and 
mass imbalances a critical interaction U$_{c}$ is required for realizing the 
uniform (q=0) superfluid state. 
In order to verify whether the requirement of U$_{c}$ is an artifact of the finite 
size lattice we have carried out the ground state as well as the finite temperature
calculations at different system sizes. Fig.20 shows the mean field ground state 
at U=3t$_{L}$ for different system sizes. We observe that for any system size
the superfluid pairing field amplitude is finite upto a population imbalance of h$_{c} \sim 0.5$. 
For h $>$ h$_{c}$ the system is a partially polarized Fermi liquid (PPFL). 
The figure shows that the regimes of various phases are stable against the choice 
of the system sizes and one can thus rule out the possibility of the finite size 
effect in the results presented in this paper. In order to validate the robustness
of our finite temperature results against the system size effects we have further 
calculated (not shown here) the BCS-BEC crossover at various system sizes. We 
observe no appreciable effect of the system size on the BCS-BEC crossover behavior
of the system.
%%%%%%%%%%%%%%%%%%%%%%%%%%%%%%%%%%%%%%%%%%%%%%%%%%%%%%%
\begin{figure}
\begin{center}
\includegraphics[height=6cm,width=6cm,angle=0]{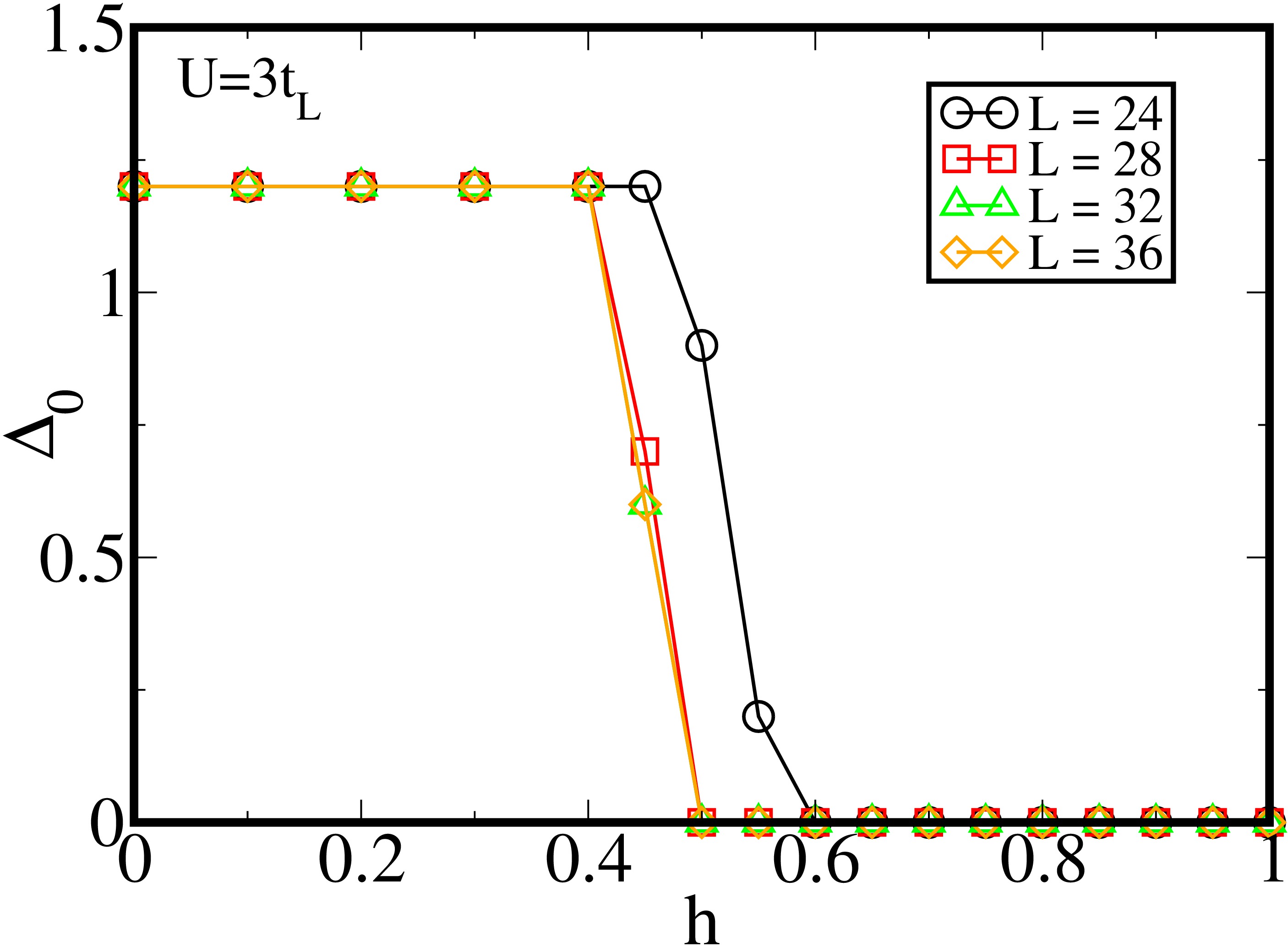}
\end{center}
\caption{Color online: Variation of the pairing field amplitude with population 
imbalance at U = 3t$_{L}$ at the ground state for different system sizes.
Notice that for sufficiently large systems the critical population imbalance 
marking the phase boundaries are independent of the system size.}
\end{figure}
%%%%%%%%%%%%%%%%%%%%%%%%%%%%%%%%%%%%%%%%%%%%%%%%%%%%%%%

%%%%%%%%%%%%%%%%%%%%%%%%%%%%%%%%%%%%%%%%%%%%%%%%%%%%%%%%%%%%%                                             
\begin{figure}
\begin{center}
\includegraphics[height=7cm,width=8cm,angle=0]{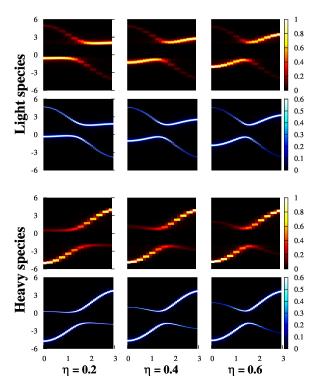}
\end{center}
\caption{Color online:Dispersion spectrum of the two species in
the BP regime (h=0.6t$_{L}$) obtained through the BdG calculation
(top panels) in comparison to those obtained through Green's function
formalism (bottom panels) at different mass imbalance ratio $\eta$.}
\end{figure}
%%%%%%%%%%%%%%%%%%%%%%%%%%%%%%%%%%%%%%%%%%%%%%%%%%%%%%%%%%%%%%%%%%%%    

\subsection{Green's function formalism}

The effect of interplay of population and mass imbalance on the quasiparticle 
properties constitute one of the key results of this work. It was observed 
that the single particle density of states (DOS) deviate significantly
 from the BCS prediction 
and reveals additional subgap and supergap features.
At the ground state a low order approximation of the 
Green's function of the electron can be set up, which can capture the quasiparticle 
behavior. The scheme is found to give fairly accurate 
results over a large $\Delta_{0}-\eta-h$ parameter space. 
The species resolved Green's function can be approximated as,

\begin{eqnarray}
 G_{LL}({\bf k},i\omega_{n}) & = & \frac{1}
 {i\omega_{n}-(\epsilon({\bf k}) + \mu_{L}) - \Sigma_{LL}
 ({\bf k},i\omega_{n})} \nonumber 
\cr G_{HH}({\bf k},i\omega_{n}) & = & \frac{1}
 {i\omega_{n}-(\epsilon({\bf k}) + \mu_{H}) - \Sigma_{HH}
 ({\bf k},i\omega_{n})} \nonumber 
\end{eqnarray}
where,
\begin{eqnarray}
 \Sigma_{LL}({\bf k},i\omega_{n}) & = & \frac{\mid \Delta_{0}\mid^{2}}{4}
 [\frac{1}{(\omega + \epsilon({\bf -k-Q}) - \mu_{H})} + 
 \nonumber \\ &&\frac{1}
 {(\omega + \epsilon({\bf -k+Q}) - \mu_{H})}] \nonumber
 \cr \Sigma_{HH}({\bf k},i\omega_{n}) & = & \frac{\mid \Delta_{0}\mid^{2}}{4}
 [\frac{1}{(\omega + \epsilon({\bf -k-Q}) - \mu_{L})} + \nonumber \\ && \frac{1}
 {(\omega + \epsilon({\bf -k+Q}) - \mu_{L})}] \nonumber 
\end{eqnarray}
with,$\epsilon_{ii}({\bf k}) = -2t_{ii}(cos(k_{x}+cos(k_{y})))$  
where, $ii = L, H$.
From the above expressions one can extract the spectral function
as, $A_{LL}({\bf k},\omega) = -(1/\pi)Im G_{LL}
({\bf k},\omega+i\delta)\mid_{\delta \rightarrow 0}$. Similar expression can 
be obtained for $A_{HH}({\bf k},\omega)$.

In Fig.21 and Fig.22 we have compared our results obtained through 
Monte carlo simulations with those obtained through the Green's 
function formalism at h=0.6t$_{L}$ and h=1.0t$_{L}$, representative
of the BP and LO phases, respectively. We have shown the spectral 
function (A$_{\sigma}$(${\bf k}, \omega$)) maps at different mass imbalance 
ratio $\eta$ for the light and heavy fermion species. The agreement between 
the results obtained through the two techniques is fairly good and 
along with capturing the species dependent behavior of the dispersion 
spectra the technique also reproduces the multi branched dispersion 
spectra for the LO state.  
The agreement justifies our choice of the Green's function formalism to 
access quasiparticle behavior at large system sizes, at the ground state. 
%%%%%%%%%%%%%%%%%%%%%%%%%%%%%%%%%%%%%%%%%%%%%%%%%%%%%%%%%%%%%%%%%%%%                                      
\begin{figure}
\begin{center}
\includegraphics[height=7cm,width=8cm,angle=0]{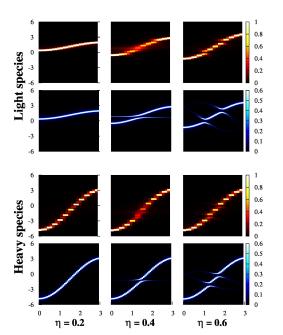}
\end{center}
\caption{Color online:Dispersion spectrum of the two species in
the LO regime (h=t$_{L}$) obtained through the BdG calculation
(top panels) in comparison to those obtained through Green's function
formalism (bottom panels) at different mass imbalance ratio $\eta$.}
\end{figure}
%%%%%%%%%%%%%%%%%%%%%%%%%%%%%%%%%%%%%%%%%%%%%%%%%%%%%%%%%%%%%   
\section{Conclusions}

In conclusion, we have investigated the BCS-BEC crossover of mass imbalanced 
Fermi-Fermi mixture within the framework of a two-dimensional lattice fermion 
model. We have mapped out the thermal phase diagram in the $\eta$-T plane and 
have shown how the thermal scales are suppressed by the imbalance in mass.
Further, investigation of the quasiparticle behavior revealed that unlike the 
balanced superfluid, the Fermi-Fermi mixture comprises of two pseudogap regimes 
as PG-I, in which the single particle excitation spectra of both the species are 
pseudogapped and PG-II, where only the light species is pseudogapped. We have 
further investigated the interplay of population imbalance in such Fermi-Fermi 
mixtures and have shown that at a fixed imbalance in population uniform
superfluidity is realized only beyond a critical U$_{c}$ unlike the balanced 
superfluid. Moreover, it was shown that a modulated LO superfluid state gives 
rise to a nodal superfluid gap in spite of an s-wave pairing field symmetry.
We have made quantitative predictions of the thermal scales 
pertaining to the 6$_{Li}$-40$_{K}$ mixture and have suggested that 
experimental techniques such as rf and momentum resolved photoemission 
spectroscopy can be used to probe the PG-I regime upto T$\sim$58nK and PG-II
regime upto T$>$108nK, in this mixture, close to the unitarity. While the 
T$_{c}$ is strongly suppressed in this mixture, signatures of short
range pair correlation survive upto significantly higher temperatures. 
We believe that our results can serve as suitable benchmarks for the 
experimental observations of 6$_{Li}$-40$_{K}$ mixture. 

\section{Acknowledgement}
The author gratefully acknowledges Prof. Pinaki Majumdar for 
the insightful comments on the manuscript. The HPC cluster 
facility of HRI, Allahabad is duly acknowledged. 
A part of this work was carried out at IMSc, Chennai,
 India and the author acknowledges the hospitality provided during the 
visit.

\bibliographystyle{apsrev4-1}
\bibliography{mass.bib}

\end{document}